# Interfaces Govern Structure of Angstrom-Scale Confined Water

Yongkang Wang,[1,2#] Fujie Tang,[3,4#] Xiaoqing Yu,[1] Kuo-Yang Chiang,[1] Chun-Chieh Yu,[1] Tatsuhiko Ohto,[5] Yunfei Chen,[2] Yuki Nagata,[1*] Mischa Bonn[1*]

[1] Max Planck Institute for Polymer Research, Ackermannweg 10, 55128 Mainz, Germany.
[2] School of Mechanical Engineering, Southeast University, 211189 Nanjing, China.
[3] Pen-Tung Sah Institute of Micro-Nano Science and Technology, Xiamen University, 361005 Xiamen, China.
[4] Laboratory of AI for Electrochemistry (AI4EC), IKKEM, 361005 Xiamen, China
[5] Graduate School of Engineering, Nagoya University, Nagoya 464-8603, Japan.
[#] Yongkang Wang and Fujie Tang contributed equally to this work.
[*] Correspondence to: nagata@mpip-mainz.mpg.de, bonn@mpip-mainz.mpg.de

## Summary

Water plays a crucial role in geological, biological, and technological processes. Nanoscale water confinement occurs in many of these settings, including sedimentary rocks, water channel proteins, and applications like desalination and water purification membranes. The structure and properties of water in nanoconfinement can differ significantly from bulk water, exhibiting, for instance, modified hydrogen bonds, dielectric constant, and phase transitions. Despite the importance of strongly nanoconfined water, experimentally elucidating the nanoconfinement effects on water, such as its orientation and hydrogen bond (H-bond) network, has remained challenging. Here, we study two-dimensionally nanoconfined aqueous electrolyte solutions with tunable confinement from nanoscale to angstrom-scale sandwiched between a graphene sheet and $CaF_2$ achieved by capillary condensation. We employ heterodyne-detection sum-frequency generation (HD-SFG) spectroscopy, a surface-specific vibrational spectroscopy capable of directly and selectively probing water orientation and H-bond environment at interfaces and under confinement. Remarkably, the vibrational spectra of the nanoconfined water can be described quantitatively by the sum of the individual water surface signals from the $CaF_2$/water and water/graphene interfaces until the confinement reduces to angstrom-scale (< ~8 Å). Machine-learning-accelerated *ab initio* molecular dynamics simulations confirm our experimental observation. These results manifest that interfacial, rather than nanoconfinement effects, dominate the water structure until angstrom-level confinement.

# Main Text

Water under nanoconfinement is the subject of increasing focus,[1–6] because nanoconfinement has the potential to modify water properties from bulk water[7–9], and water in nanoconfinement provides a unique platform for chemical reactions.[10–12] For example, water in nanoconfinement exhibits anomalously low dielectric constant,[8] engages in weakened/strengthened hydrogen bonding,[13,14] and undergoes structural phase transition.[4,5,15] Water and ions under nanoconfinement show extraordinary transport properties.[16,17] Water molecules are more reactive in nanoconfinement than in bulk, spawning the realm of "chemistry in confinement".[10–12,18] The anomalous water behavior in nanoconfinement is usually ascribed to the "nanoconfinement effects". In many studies of properties of confined water, deviations from bulk theory expectations already appear at modest (~10 nm) confinement.[8] An urgent and important question is whether these effects are due to the "nanoconfinement effects" on the water itself, or rather due to the "interfacial effects",[19,20] which states that the properties of water under nanoconfinement are altered by water in contact with two interfaces. Indeed, the structure of interfacial water also differs from bulk water, with a truncated hydrogen bond network and ordered structure due to the interaction with the surface.[21,22] A fundamental question regarding the "nanoconfinement effects" is at what level of confinement the "nanoconfinement effects" on the water start to appear so that the structure of water under nanoconfinement deviates from the cooperative "interfacial effects".

To address the question and differentiate between the "interfacial effects" and "nanoconfinement effects" on water, molecular-level insights into nanoconfined water and interfacial water, such as their orientation and H-bond structure, are necessary. To this end, heterodyne-detection sum-frequency generation (HD-SFG) spectroscopy is uniquely suited, owing to three advantages: molecular specificity (in particular, sensitivity to hydrogen bonding strength and molecular orientation); interface and confinement specificity; and additivity of the molecular water response. In particular, the third property allows us to distinguish the additivity of two interfacial effects from true confinement effects. In an SFG experiment, infrared and visible laser fields are mixed to generate the sum frequency of those two fields. The signal is enhanced when the infrared (IR) frequency resonates with the molecular vibration, providing specificity to molecular structure.[23] Furthermore, HD-SFG spectroscopy provides complex-valued $\chi^{(2)}$ spectra and the sign of the imaginary part of the $\chi^{(2)}$ spectrum (Im$\chi^{(2)}$) reflects the absolute orientation of water molecules (*up-/down*-orientation).[24] SFG signals are non-zero only when Centro-symmetry is broken, such as at water interface[21,25,26] or nanoconfinement-induced alignment of water.[27] Signals from the bulk water are naturally excluded due to the SFG selection rule.[28] As such, it can selectively distinguish water experiencing the "nanoconfinement effects" and the "interfacial effects". Furthermore, an HD-SFG signal ($\chi^{(2)}$) is additive and thus allows us to quantitatively disentangle the contributions from the two interfaces and the nanoconfinement in the nanoconfined system. With these advantages, one can directly compare the structure of the nanoconfined water with the structure of the interfacial water at an edge of the bulk water, opening a path to distinguish the difference between the "interfacial effects" and the "nanoconfinement effects" on water.

Nevertheless, applying HD-SFG spectroscopy to the realm of nanoconfined liquids has been challenging owing to 1) the length mismatch between nanofluidic devices and typical laser spots: for extreme (sub-nanometer) confinement, devices are dimensioned below a few hundred nanometers[8,16,29] while the SFG probes a spot of ~100 μm; and 2) the challenge of the IR beam to reach the confined region. We overcome these two challenges by fabricating a centimeter-sized nanoconfined water system with tunable confinement from nanoscale to angstrom-scale sandwiched between a flat calcium fluoride (CaF$_2$) substrate and a graphene sheet. The tunable two-dimensional nanoconfinement is achieved by capillary condensation[30,31] between the hydrophilic CaF$_2$ substrate and the graphene sheet. Furthermore, both the CaF$_2$ substrate and the graphene sheet allow the IR beam to reach the confined surface. By applying HD-SFG spectroscopy to the nanoconfined water system with tunable confinement and comparing the nanoconfined water response to the response of water at the CaF$_2$/bulk water interface and bulk water/graphene interface, we identify the transition, at angstrom-scale confinement, from water dominated by "interfacial effects" to that dominated by "nanoconfinement effects".

We prepared the nanoconfined water sample sandwiched by a graphene sheet and a hydrophilic $CaF_2$ substrate by trapping water molecules with $Li^+$ and $Cl^-$ ions under controlled relative humidity (RH) condition (Methods and Supplementary Methods).[2,6,30] Using this method, the confined region is composed of the aqueous LiCl solution with a LiCl concentration in the ~few molar ranges (Supplementary Discussion S1). We ensured that the $CaF_2$ substrate is atomically smooth, with a root mean square (RMS) surface roughness close to that of graphite. The flatness of the $CaF_2$ substrate and the homogeneity of the confined water sample are discussed in Supplementary Discussions S2 and S3. The high quality of the nanoconfined water sample's graphene sheet was verified using Raman microscopy. The Raman data shows the absence of the D-band (~1350 $cm^{-1}$), indicating that the graphene sheet has no obvious defect.[32] The 2D-band and G-band intensity ratio of 3 confirms that the graphene sheet comprises a monolayer (see Supplementary Discussion S4 for more details).

Subsequently, we measured the thickness of the nanoconfined water using atomic force microscopy (AFM) at a controlled RH ~25% (see Methods). To ensure that the probed sample region was the same for AFM and HD-SFG spectroscopy, a gold structure was used to mark a sample region of approximately 250×200 μm² at the edge of the graphene sample (Figs. 1a and 1b, see Supplementary Discussion S3 for more details). An AFM image probing the edge of the graphene sheet is presented in Fig. 1c, while a typical height profile indicated by the white dashed line in the AFM image is shown in Fig. 1d. The edge of the graphene sheet is folded over, producing a sharp height peak. The height difference between the region where the graphene sheet covers and the region where the graphene is absent is 11.5 ± 0.6 Å. By using the estimated thickness of the exclusive volume of the graphene sheet of 3.3 Å from the *ab initio* molecular dynamics (AIMD) simulation (see Fig. S4 in Supplementary Methods), we conclude that the thickness of the confined water in our samples is 8.2 ± 0.6 Å. A schematic of the composition of the nanoconfined water sample is shown in Fig. 1e. Furthermore, the AIMD simulation indicates the ~8 Å confined water is composed of three water layers. The confined three-layer water with its thickness of ~8 Å at an RH of ~25% can be accounted for by the capillary condensation under hydrophilic confinement (see Supplementary Discussion S5 for more details), consistent with previous studies.[30,33]

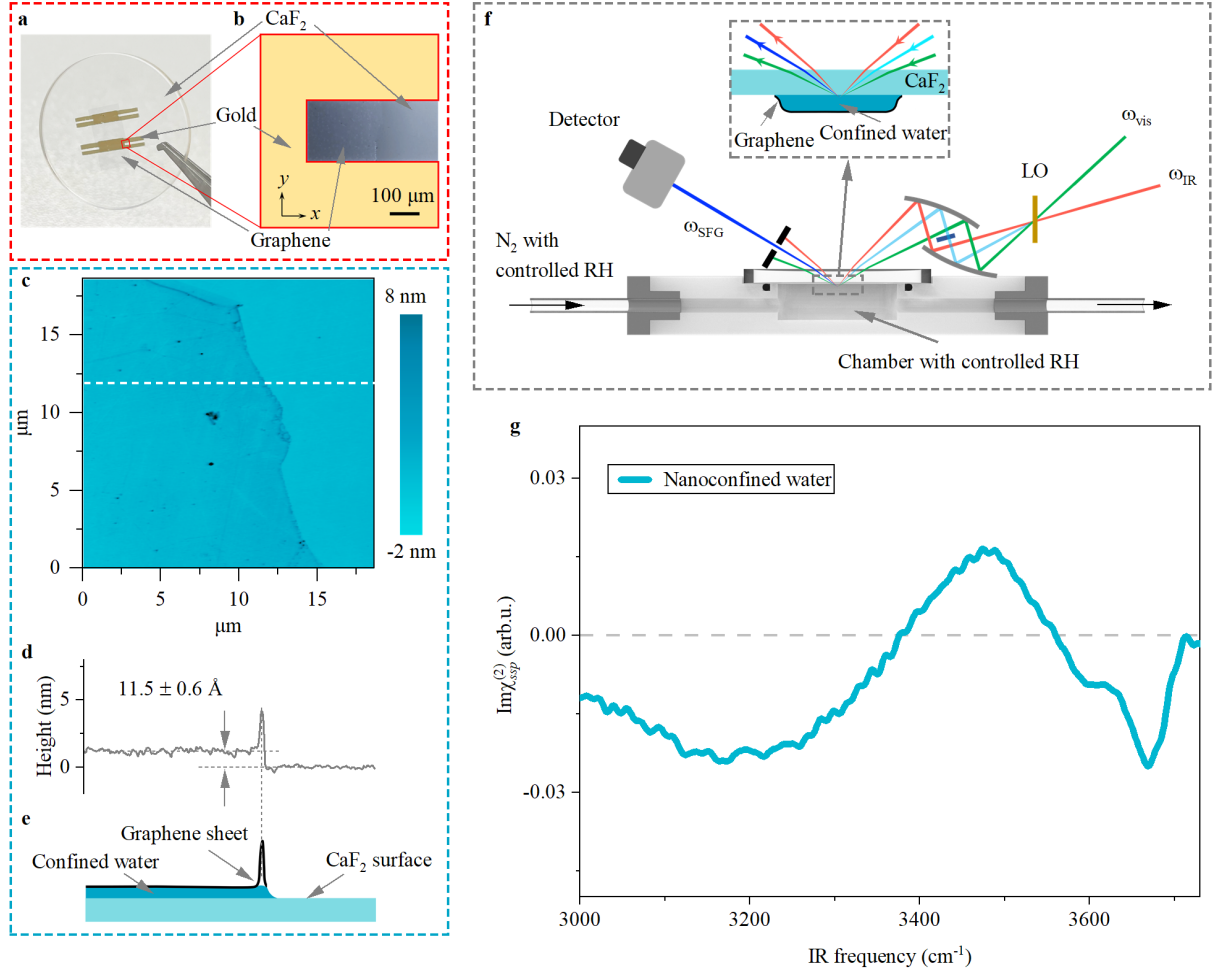

**Fig. 1 | HD-SFG spectroscopy of nanoconfined water. a**. An optical image of the nanoconfined water sample surrounded with gold film. The grey shade area indicates the region covered by graphene. **b**. A close-up optical image of the nanoconfined water sample. **c**. AFM image of the edge of the graphene sheet of the nanoconfined sample. **d**. The height profile along the white dashed line in (**c**), crossing the graphene wrinkles. The error bar of the sample height was a 95% confidence interval (CI) of data measured on five different samples. **e**. A schematic of the composition of the nanoconfined sample. **f**. Illustration of HD-SFG experiment. **g**. The $\mathrm{Im}\chi^{(2)}_{ssp}$ spectrum of the nanoconfined water sandwiched between the graphene sheet and the $CaF_2$ substrate. The grey dashed line serves as a zero line.

We carried out HD-SFG measurement within the marked region on the nanoconfined water sample at the *ssp* polarization combination with the three letters indicating the polarizations of the SFG, visible, and infrared light fields, respectively (see Fig. 1f). The $\mathrm{Im}\chi^{(2)}_{ssp}$ spectrum for the confined three-layer water system measured at RH ~25% is displayed in Fig. 1g. This spectrum exhibits negative ~3670 cm$^{-1}$, positive ~3460 cm$^{-1}$, and negative ~3200 cm$^{-1}$ peaks. A positive (negative) peak in the O-H stretch $\mathrm{Im}\chi^{(2)}_{ssp}$ spectrum corresponds to the *up*- (*down*-)oriented O-H group to the CaF$_2$ substrate, with *down* pointing towards the graphene sheet.[21,34–36] Furthermore, low (high) O-H stretch frequencies peak indicate a strong (weak) H-bond strength of the O-H group.[37] The mixture of the positive and negative peaks spanning over the 3000-3700 cm$^{-1}$ frequency region indicates that these O-H groups with different H-bond strengths are oriented differently.

The interpretation of the SFG spectrum of the nanoconfined water is complicated because two interfaces are probed simultaneously.[27] To uncover the origin of the SFG features for the nanoconfined water system, we compared the SFG spectrum of the nanoconfined water with the spectra measured at the bulk LiCl solution/suspended graphene interface and the CaF$_2$/bulk LiCl solution interface (see Supplementary Methods). For brevity, we refer to the LiCl solution as 'water' afterward. The quantitative comparison of the water SFG signals from these different samples requires conversion of the measured $\mathrm{Im}\chi^{(2)}_{ssp}$ into $\mathrm{Im}\chi^{(2)}_{yyz}$ spectra by correcting for Fresnel factors and reflectivity using the three-layer dielectric model[38,39] (see Supplementary Discussion S6 and S7). The validity of this analysis will be justified below. The inferred $\mathrm{Im}\chi^{(2)}_{yyz}$ spectra for the nanoconfined water system, the water/graphene interface, and the CaF$_2$/water interface ($\mathrm{Im}\chi^{(2)}_{yyz,\,\mathrm{confined}}$, $\mathrm{Im}\chi^{(2)}_{yyz,\mathrm{W/G}}$, and $\mathrm{Im}\chi^{(2)}_{yyz,\mathrm{CaF_2/W}}$) are displayed in Fig. 2a, b, and c, respectively.

The $\mathrm{Im}\chi^{(2)}_{yyz,\,\mathrm{confined}}$ and $\mathrm{Im}\chi^{(2)}_{yyz,\mathrm{W/G}}$ spectra both contain the negative 3670 cm$^{-1}$ peak with almost the same amplitude. This peak arises from a dangling O-H group pointing *down* to the graphene sheet[21,34], weakly interacting with the π-orbital of the graphene sheet without forming H-bonds with other water molecules.[21,34,40] In contrast, the negative hydrogen-bonded (H-bonded) O-H peak at ~3200 cm$^{-1}$ in the $\mathrm{Im}\chi^{(2)}_{yyz,\,\mathrm{confined}}$ spectrum is missing in the $\mathrm{Im}\chi^{(2)}_{yyz,\mathrm{W/G}}$ spectrum. Furthermore, the peak amplitude of the positive H-bonded O-H peak at ~3460 cm$^{-1}$ in the $\mathrm{Im}\chi^{(2)}_{yyz,\,\mathrm{confined}}$ spectrum is much weaker than that in the $\mathrm{Im}\chi^{(2)}_{yyz,\mathrm{W/G}}$

spectrum. Such discrepancies can be attributed to the impact of the CaF$_2$ surface on the SFG spectrum of the nanoconfined water. In fact, the Im$\chi^{(2)}_{yyz,\text{CaF}_2/\text{W}}$ spectrum shows the negative H-bonded O-H peak at ~3350 cm$^{-1}$, explaining the appearance of the negative 3200 cm$^{-1}$ peak in the Im$\chi^{(2)}_{yyz,\text{confined}}$ spectrum. This negative 3350 cm$^{-1}$ peak in the Im$\chi^{(2)}_{yyz,\text{CaF}_2/\text{W}}$ spectrum arises from the interfacial water pointing away from the positively charged CaF$_2$ surface.[36]

Remarkably, the SFG signal for the ~8 Å confined three-layer water can be quantitatively accounted for by combining the water/graphene interface and CaF$_2$/water interface signals, i.e.:

$$\chi^{(2)}_{yyz,\text{confined}} = \chi^{(2)}_{yyz,\text{W/G}} + \chi^{(2)}_{yyz,\text{CaF}_2/\text{W}}. \tag{1}$$

The comparison between the Im$\chi^{(2)}_{yyz,\text{confined}}$ spectrum and the summed spectrum of the two Im$\chi^{(2)}_{yyz,\text{W/G}}$ and Im$\chi^{(2)}_{yyz,\text{CaF}_2/\text{W}}$ spectra is shown in Fig. 2a. The two spectra overlap within the experimental uncertainty, manifesting that Eq. (1) holds (see also the comparison of the real part of the two spectra in Supplementary Discussion S8). Such consistency suggests that the confinement itself induces no effect on the water organization (orientations and H-bond strength) for the confined three-layer water, and the organization of the confined three-layer water is governed by the superposition of the two independent interfaces.

Note that $\chi^{(2)}_{yyz,\text{W/G}}$ spectrum is insensitive to the ion concentration, while $\chi^{(2)}_{yyz,\text{CaF}_2/\text{W}}$ spectrum at the charged CaF$_2$/water interface is sensitive due to the so-called $\chi^{(3)}$ contribution.[41–43] As a result, the ion concentration of the nanoconfined LiCl solution may affect the validity of Eq. (1). Our HD-SFG spectral analysis indicates the ~8 Å confined LiCl solution is in the ~few molar ranges (~2-3 M) and within the range of conceivable LiCl concentrations (2-6 M), Eq. (1) holds (see Supplementary Discussion S1).

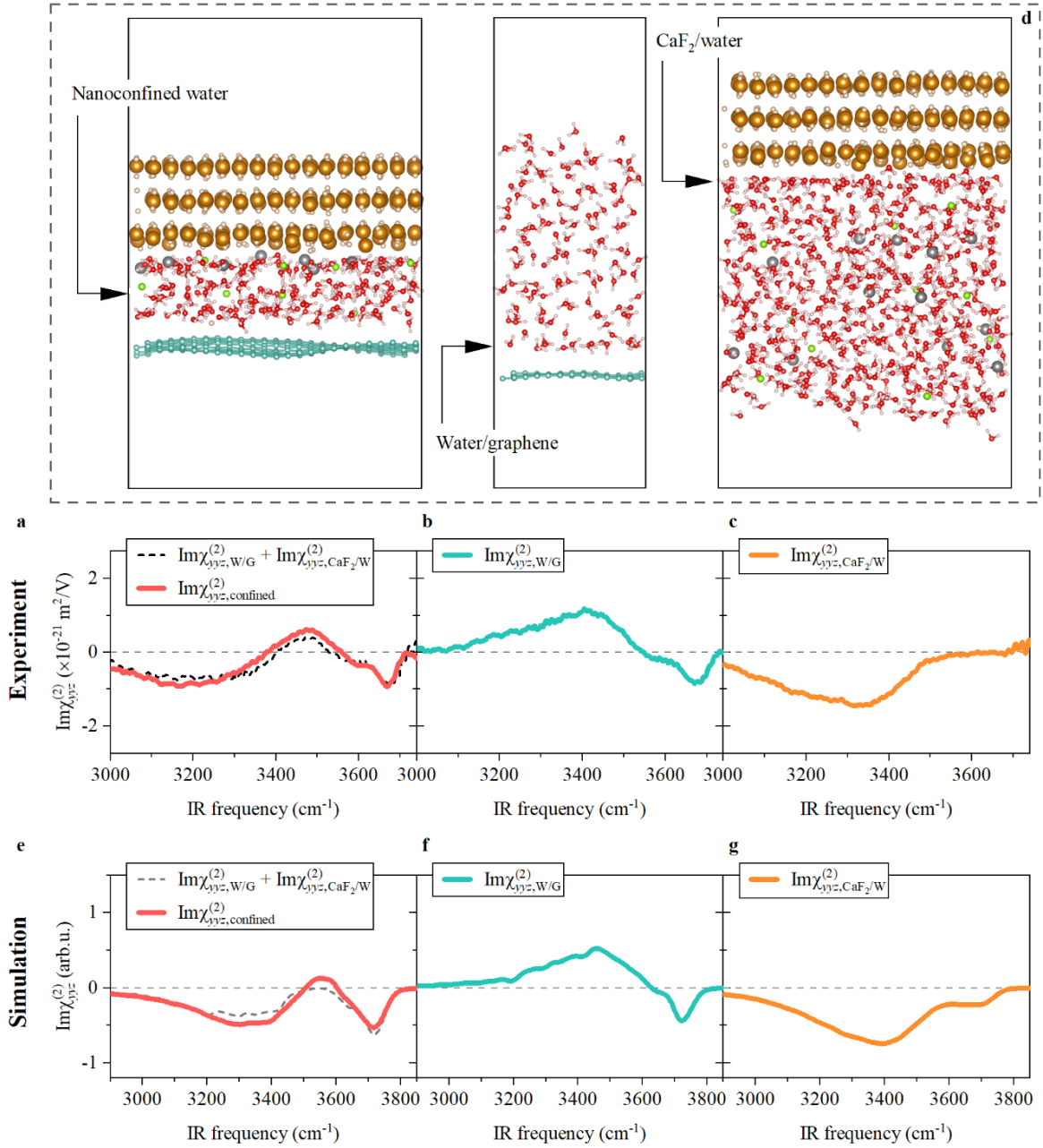

**Fig. 2 | Nanoconfined water organization is governed by interfacial effects. a**. Experimental $\mathrm{Im}\chi^{(2)}_{yyz}$ spectra of nanoconfined three-layer water (red solid line). The sum of the water/graphene and $CaF_2$/water signals obtained from HD-SFG experiments is also shown for comparison (black dashed line). **b**, **c**. Experimental $\mathrm{Im}\chi^{(2)}_{yyz}$ spectra at the (**b**) bulk water/graphene interface and (**c**) $CaF_2$/bulk water interface. We used a 2 M LiCl aqueous solution for the HD-SFG experiments shown in (**b**) and (**c**). **d**. Snapshots of the nanoconfined three-layer water (MLFF-MD), water/graphene interface (AIMD), and $CaF_2$/water interface (MLFF-MD) obtained from the MD simulations. The yellow, light yellow, red, light pink, green,

grey, and cyan spheres indicate the Ca, F, O, H, Cl, Li, and C atoms, respectively. **e**, **f**, **g**. Theoretical $\mathrm{Im}\chi_{yyz}^{(2)}$ spectra of (**e**) the nanoconfined three-layer water (red solid line); (**f**) the water/graphene interface; and (**g**) the CaF$_2$/water interface. The sum of the water/graphene (**e**) and CaF$_2$/water (**f**) signals is shown in (**e**) for comparison (black dashed line). Dashed lines in (**a-c**) and (**e-g**) serve as zero lines.

To convert the recorded $\mathrm{Im}\chi_{ssp}^{(2)}$ response from the three different samples into experimental $\mathrm{Im}\chi_{yyz}^{(2)}$ spectra requires Fresnel factor and reflectivity corrections.[38,39] To do so, we employed the three-layer dielectric model, but the choice of model affects the inferred $\mathrm{Im}\chi_{yyz}^{(2)}$.[44,45] Thus, it is essential to independently validate the $\mathrm{Im}\chi_{yyz}^{(2)}$ spectra. To this end, we carried out the SFG spectra simulation for the confined three-layer water (LiCl solution) sample and CaF$_2$/bulk LiCl solution sample with machine learning force field MD (MLFF-MD) simulations, as well as the bulk water/graphene sample with AIMD simulations (see Supplementary Methods). The simulation allows us to directly access the $\mathrm{Im}\chi_{yyz}^{(2)}$ spectra so the validity of Eq. (1) can be examined critically.[39,44]

The snapshots of the MLFF-MD/AIMD simulations are shown in Fig. 2d, while the simulated $\mathrm{Im}\chi_{yyz}^{(2)}$ spectra of the confined three-layer water system, the water/graphene interface, and the CaF$_2$/water interface are displayed in Fig. 2e, f, and g, respectively. The lineshapes of the simulated $\mathrm{Im}\chi_{yyz}^{(2)}$ spectra for all these three systems closely resemble the experimental data. Furthermore, the relative peak amplitudes of the SFG spectra for these three systems also agree with the experimental data. Notably, the amplitude of the positive peak at ~3450 cm$^{-1}$ at the water/graphene interface is close to that of the negative peak at ~3400 cm$^{-1}$ at the CaF$_2$/water interface. Nevertheless, due to the slightly different peak frequencies of the positive peak and the negative peak, the sum of the two SFG spectra creates a negative-positive feature in the 3000-3600 cm$^{-1}$ region (Fig. 2e). Eventually, the summed SFG data agrees well with the SFG data for the nanoconfined water system, meaning that Eq. (1) holds also for the simulated SFG spectra. This demonstrates that even with its thickness of three-layer water, the nanoconfinement itself induces no unique water organization. The organization of the confined three-layer water is governed by the interfacial water, manifesting as a simple superposition of the two independent interface systems. Similar interfacial structure of water for the two bulk water samples and the nanoconfined water sample is also apparent from the joint probability plot of the two O-H bonds and water dipole orientation analysis,[46] presented in the Supplementary Information (Supplementary Discussion S9). Furthermore, the Fresnel factor corrections and reflectivity corrections we employed in the experiment are sufficient to provide accurate amplitude calibration for the measured SFG data.

The question arises at what level of confinement the "nanoconfinement effects" start to appear and modify the water structure so the structure of the water under nanoconfinement deviates from the cooperative "interfacial effects". To address this question, we measured the $\text{Im}\chi^{(2)}_{yyz}$ spectra on the nanoconfined water sample while tuning the thickness of the nanoconfined water, achieved by changing the RH in the sample cell (see Supplementary Methods).[30,31] The data are shown in Fig. 3. As expected, the $\text{Im}\chi^{(2)}_{yyz,\text{confined}}$ spectrum remains unaffected when increasing the water thickness above 8 Å, but it changes substantially when reducing the thickness from ~8 Å to ~5 Å (see AFM data in Supplementary Discussion S5). The negative 3200 cm$^{-1}$ peaks and the positive 3460 cm$^{-1}$ peak decrease dramatically, and the 3670 cm$^{-1}$ peak nearly disappears. Such spectral change is more significant with further decreasing the water thickness down to ~3 Å. Clearly, Eq. (1) is no longer valid for the nanoconfinement below ~8 Å. Lowering the RH decreases the thickness of the confined water and at the same time, increases its salt concentration. We note both the $\chi^{(2)}_{yyz,\text{W/G}}$ and $\chi^{(2)}_{yyz,\text{CaF}_2/\text{W}}$ are insensitive to ion concentration at high ion concentration ($\geq 2$ M), the observed abrupt change of the $\text{Im}\chi^{(2)}_{yyz}$ spectra upon decreasing the RH cannot be simply explained by the increase of ion concentration in confinement (See Supplementary Discussion S1 and S10 for more details). Rather, these substantial spectral changes indicate that the "nanoconfinement effects" start to appear and gradually dominate. The decrease of the two H-bonded O-H peaks and the disappearance of the 3670 cm$^{-1}$ dangling O-H peak imply that the <8 Å confinement forces the confined water to lie flat with O-H groups parallel to the graphene and the CaF$_2$ substrate, consistent with recent theoretical predictions.[4] The parallel-orientated O-H groups yield zero SFG signals so the peak intensities decrease. The overall shift of the spectral intensity to higher frequencies denotes that the H-bond network is weakened substantially due to the enhanced confinement. The HD-SFG measurements on different thicknesses of the nanoconfined water demonstrate that the "nanoconfinement effects" on the water appear and dominate when the confinement reaches the Angstrom scale (< ~8 Å, see Supplementary Discussion S11-14 for more simulation data).

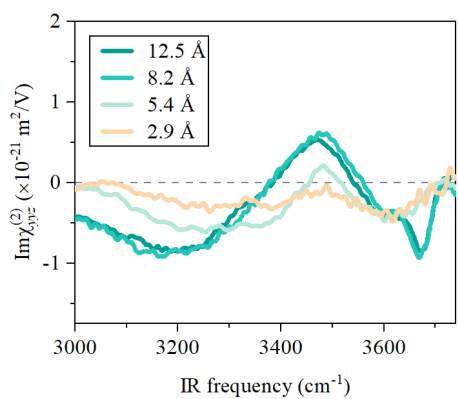

**Fig. 3 | Nanoconfinement effect on water.** Experimental $\mathrm{Im}\chi^{(2)}_{yyz}$ spectra of the nanoconfined water of different thicknesses. The dashed line serves as a zero line.

Our work addresses the fundamental question related to the "nanoconfinement effect" on water. "Interfacial effects" rather than "nanoconfinement effects" govern the molecular structure of nanoscale confined water until the confinement reduces to angstrom-scale (< ~8 Å). The revelation of the crucial role of "interfacial effects" on nanoscale confined water (≥ 8 Å) could serve to connect the realms of "interfacial chemistry" and "chemistry in confinement", and offer insight into explaining anomalous phenomena observed in nanoscale confinement, such as the anomalously low dielectric constant of nanoconfined water[8] and the anomalous ionic transport in nanochannels.[17] Moreover, our findings indicate that for engineering and improving technologies requiring nanoscale confinement, focusing on engineering interfacial properties of the confining materials may suffice when confinement does not occur below ~1 nm. Furthermore, the revelation of the "nanoconfinement effects" on water not occurring until angstrom-scale confinement (< 8 Å) illustrates the necessity to create angstrom-scale confinement to induce 'true' confinement-induced phenomena. Finally, our work demonstrates that HD-SFG spectroscopy is powerful and highly sensitive for investigating confined water's orientation and H-bond network structure even for confinement down to angstrom-scale. The findings and experimental capability established here raise the exciting prospect of future work regarding the probe of molecular details of water and ionic transport in extreme confinement.

## Methods

***Sample Preparation.*** The nanoconfined water samples were prepared by using the wet transfer technique[2,47] to enclose water between a graphene sheet and a $CaF_2$ substrate (Fig. S1, see Supplementary Methods in the Supplementary Information for more details). In brief, we used a commercial CVD-grown large-area monolayer graphene sheet on copper foil (Grolltex Inc). The graphene sheet was exposed to LiCl solution and was then transferred onto the $CaF_2$ substrate. A thin film of LiCl solution was trapped between the graphene and the substrate due to the hydrophilic nature of the $CaF_2$ substrate. Upon drying for ~12 hours in air with an RH of ~25%, nanoconfined water samples were obtained as a result of capillary condensation[30] (see Supplementary Discussion S5 for more details). The preparation of the suspended graphene on the water surface was similar to Refs.[40,48] and was detailed in our recent work.[49] (Fig. S2, see Supplementary Methods for more details).

***HD-SFG Measurement.*** HD-SFG measurements were performed on an HD-SFG setup in a non-collinear beam geometry with a Ti: Sapphire regenerative amplifier laser system. A detailed description can be found in Refs.[22,36]. HD-SFG spectra were measured in a dried air atmosphere to avoid spectral distortion due to water vapor. The sample cell was purged with $N_2$ of different RHs during the measurement to tune the thickness of the nanoconfined water (see Supplementary Methods). We checked the sample height with a displacement sensor (CL-3000, Keyence with a resolution of ~0.1 μm). The IR, visible, and LO beams were directed at the sample at incidence angles of 33°, 39°, and 37.6°, respectively. The measurements were performed at the *ssp* polarization combination, where *ssp* denotes *s*-polarized SFG, *s*-polarized visible, and *p*-polarized IR beams. More details about the sample preparation, sample cell,

Raman measurement, HD-SFG measurement, and data analysis can be found in Sections 1-8 in the Supplementary Method of the Supplementary Information.

*AFM Measurement.* The surface morphology of the graphene samples was measured using an atomic force microscope (AFM, Bruker, JPK) working in the noncontact mode. We used a silicon cantilever (OLYMPUS OMCL-AC 160, f = 243 kHz, k = 25 N m$^{-1}$) for the measurement, and the measurement was conducted in the air of different RHs. We control the RH inside the AFM chamber by purging with air of different RHs.

*SFG Spectral Simulation.* We performed the MD simulation at the *ab initio* level of theory with the help of machine learning and we computed the SFG spectra based on the obtained MD trajectories. We performed the MD simulations at the water/graphene interface and the CaF$_2$/aqueous LiCl solution interface as well as the confined water systems using varying thicknesses of the confined water. The AIMD simulations were carried out by using the CP2K code.[50] From the AIMD data, we created the force field through machine learning technique using the DeepMD-kit package.[51] The cell size of the graphene/water interface was composed of $\vec{a}$, $\vec{b}$, and $\vec{c}$, while the cell sizes of the CaF$_2$/water interface and nanoconfined systems were $2\vec{a}$, $2\vec{b}$, and $\vec{c}$, where $\vec{a} = (14.76 \text{ Å}, 0 \text{ Å}, 0 \text{ Å})$, $\vec{b} = (7.38 \text{ Å}, 12.78 \text{ Å}, 0 \text{ Å})$, and $\vec{c} = (0 \text{ Å}, 0 \text{ Å}, 50 \text{ Å})$. The MLFF-MD simulations were performed in the NVT ensemble with the target temperature of 300 K by using the LAMMPS code.[52] Using the coordinates and velocities of water molecules, we computed the SFG spectra within the surface-specific velocity-velocity correlation function formalisms.[53] Details of the simulation can be found in Sections 9-11 in the Supplementary Method of the Supplementary Information.

## Data availability

All study data are included in the main text and/or Supplementary Information.

## Notes

The authors declare no competing financial interest.

# Author contributions

Y.K.W., Y.N., and M.B. designed the study. Y.K.W. prepared the samples and performed the AFM and Raman microscope measurements. Y.K.W., X.Q.Y., K.Y.C., and C.C.Y. performed HD-SFG measurements. T.O. and Y.N. conducted the AIMD simulations. F.T. conducted the MLFF-MD simulations and computed the SFG spectra. The data analysis was done by Y.K.W. (experiment) and F.T. (simulation). Y.K.W., F.T., Y.N., and M.B. wrote the manuscript. All authors contributed to interpreting the results and refining the manuscript.

# Acknowledgments

We are grateful for the financial support from the MaxWater Initiative of the Max Planck Society. Funded by the European Union (ERC, n-AQUA, 101071937). Views and opinions expressed are however those of the author(s) only and do not necessarily reflect those of the European Union or the European Research Council Executive Agency. Neither the European Union nor the granting authority can be held responsible for them. We thank Johannes Hunger, Nikita Kavokine, and Maksim Grechko for providing insightful comments and suggestions on this work. We thank Ruediger Berger for providing support on AFM measurements. We also thank Florian Gericke, Marc-Jan van Zadel, and the technical workshop at the Max Planck Institute for Polymer Research for excellent technical support. F.T. is supported by a startup fund at Xiamen University. Part of this work used the computational resources in the IKKEM intelligent computing center.

# References

1. Verdaguer, A., Sacha, G. M., Bluhm, H. & Salmeron, M. Molecular Structure of Water at Interfaces: Wetting at the Nanometer Scale. *Chem. Rev.* **106**, 1478–1510 (2006).

2. Xu, K., Cao, P. & Heath, J. R. Graphene visualizes the first water adlayers on mica at ambient conditions. *Science* **329**, 1188–1191 (2010).

3. Giovambattista, N., Rossky, P. J. & Debenedetti, P. G. Effect of pressure on the phase behavior and structure of water confined between nanoscale hydrophobic and hydrophilic plates. *Phys. Rev. E* **73**, 041604 (2006).

4. Kapil, V., Schran, C., Zen, A., Chen, J., Pickard, C. J., & Michaelides, A. The first-principles phase diagram of monolayer nanoconfined water. *Nature* **609**, 512–516 (2022).

5. Takaiwa, D., Hatano, I., Koga, K. & Tanaka, H. Phase diagram of water in carbon nanotubes. *Proc. Natl Acad. Sci. USA* **105**, 39–43 (2008).

6. Algara-Siller, G., Lehtinen, O., Wang, F. C., Nair, R. R., Kaiser, U., Wu, H. A., ... & Grigorieva, I. V. Square ice in graphene nanocapillaries. *Nature* **519**, 443 (2015).

7. Chmiola, J. Anomalous Increase in Carbon Capacitance at Pore Sizes Less Than 1 Nanometer. *Science* **313**, 1760–1763 (2006).

8. Fumagalli, L., Esfandiar, A., Fabregas, R., Hu, S., Ares, P., Janardanan, A., ... & Geim, A. K. Anomalously low dielectric constant of confined water. *Science* **360**, 1339–1342 (2018).

9. Robin, P., Kavokine, N. & Bocquet, L. Modeling of emergent memory and voltage spiking in ionic transport through angstrom-scale slits. *Science* **373**, 687–691 (2021).

10. Lim, C. H. Y. X., Nesladek, M. & Loh, K. P. Observing High-Pressure Chemistry in Graphene Bubbles. *Angew. Chem. Int. Ed.* **53**, 215–219 (2014).

11. Beaumont, M., Jusner, P., Gierlinger, N., King, A. W., Potthast, A., Rojas, O. J., & Rosenau, T. Unique reactivity of nanoporous cellulosic materials mediated by surface-confined water. *Nat Commun* **12**, 2513 (2021).

12. Kobayashi, J., Mori, Y., Okamoto, K., Akiyama, R., Ueno, M., Kitamori, T., & Kobayashi, S. A Microfluidic Device for Conducting Gas-Liquid-Solid Hydrogenation Reactions. *Science* **304**, 1305–1308 (2004).

13. Erwan, P., Jean-Blaise, B., Patrick, J., Stéphan, R., Pascale, L., & Pascale, R. Water in Carbon Nanotubes: The Peculiar Hydrogen Bond Network Revealed by Infrared Spectroscopy. *J. Am. Chem. Soc.* **138**, 10437–10443 (2016).

14. Mallamace, F., Broccio, M., Corsaro, C., Faraone, A., Majolino, D., Venuti, V., ... & Chen, S. H. Evidence of the existence of the low-density liquid phase in supercooled, confined water. *Proc. Natl Acad. Sci. USA* **104**, 424–428 (2007).

15. Han, S., Choi, M. Y., Kumar, P. & Stanley, H. E. Phase transitions in confined water nanofilms. *Nature*

*Phys* **6**, 685–689 (2010).

16. Robin, P., Emmerich, T., Ismail, A., Niguès, A., You, Y., Nam, G. H., ... & Bocquet, L. Long-term memory and synapse-like dynamics in two-dimensional nanofluidic channels. *Science* **379**, 161–167 (2023).

17. Duan, C. & Majumdar, A. Anomalous ion transport in 2-nm hydrophilic nanochannels. *Nature Nanotech* **5**, 848–852 (2010).

18. Muñoz-Santiburcio, D. & Marx, D. Confinement-Controlled Aqueous Chemistry within Nanometric Slit Pores. *Chem. Rev.* **121**, 6293–6320 (2021).

19. Chialvo, A. A. & Vlcek, L. Can We Describe Graphene Confined Water Structures as Overlapping of Approaching Graphene–Water Interfacial Structures? *J. Phys. Chem. C* **120**, 7553–7561 (2016).

20. Bocquet, L. & Charlaix, E. Nanofluidics, from bulk to interfaces. *Chem. Soc. Rev.* **39**, 1073–1095 (2010).

21. Ohto, T., Tada, H. & Nagata, Y. Structure and dynamics of water at water–graphene and water–hexagonal boron-nitride sheet interfaces revealed by ab initio sum-frequency generation spectroscopy. *Phys. Chem. Chem. Phys.* **20**, 12979–12985 (2018).

22. Wang, Y., Seki, T., Yu, X., Yu, C. C., Chiang, K. Y., Domke, K. F., ... & Bonn, M. Chemistry governs water organization at a graphene electrode. *Nature* **615**, E1–E2 (2023).

23. Shen, Y. R. Surface properties probed by second-harmonic and sum-frequency generation. *Nature* **337**, 519–525 (1989).

24. Yamaguchi, S. & Tahara, T. Heterodyne-detected electronic sum frequency generation: "Up" versus "down" alignment of interfacial molecules. *J. Chem. Phys.* **129**, 101102 (2008).

25. Du, Q., Superfine, R., Freysz, E. & Shen, Y. R. Vibrational spectroscopy of water at the vapor/water interface. *Phys. Rev. Lett.* **70**, 2313 (1993).

26. Du, Q., Freysz, E. & Shen, Y. R. Surface Vibrational Spectroscopic Studies of Hydrogen Bonding and Hydrophobicity. *Science* **264**, 826–828 (1994).

27. Das, B., Ruiz-Barragan, S. & Marx, D. Deciphering the Properties of Nanoconfined Aqueous Solutions by Vibrational Sum Frequency Generation Spectroscopy. *J. Phys. Chem. Lett.* **14**, 1208–1213 (2023).

28. Bonn, M., Nagata, Y. & Backus, E. H. G. Molecular Structure and Dynamics of Water at the Water–Air Interface Studied with Surface-Specific Vibrational Spectroscopy. *Angew. Chem. Int. Ed.* **54**, 5560–5576 (2015).

29. Aluru, N. R., Aydin, F., Bazant, M. Z., Blankschtein, D., Brozena, A. H., de Souza, J. P., ... & Zhang, Z. Fluids and Electrolytes under Confinement in Single-Digit Nanopores. *Chem. Rev.* **123**, 2737–2831 (2023).

30. Yang, Q., Sun, P. Z., Fumagalli, L., Stebunov, Y. V., Haigh, S. J., Zhou, Z. W., ... & Geim, A. K. Capillary condensation under atomic-scale confinement. *Nature* **588**, 250–253 (2020).

31. Fisher, L. R., Gamble, R. A. & Middlehurst, J. The Kelvin equation and the capillary condensation of water. *Nature* **290**, 575–576 (1981).

32. Liang, X., Sperling, B. A., Calizo, I., Cheng, G., Hacker, C. A., Zhang, Q., ... & Richter, C. A. Toward Clean and Crackless Transfer of Graphene. *ACS Nano* **5**, 9144–9153 (2011).

33. Verdaguer, A., Weis, C., Oncins, G., Ketteler, G., Bluhm, H., & Salmeron, M. Growth and Structure of Water on SiO2 Films on Si Investigated by Kelvin Probe Microscopy and in Situ X-ray Spectroscopies. *Langmuir* **23**, 9699–9703 (2007).

34. Zhang, Y., de Aguiar, H. B., Hynes, J. T. & Laage, D. Water Structure, Dynamics, and Sum-Frequency Generation Spectra at Electrified Graphene Interfaces. *J. Phys. Chem. Lett.* **11**, 624–631 (2020).

35. Nihonyanagi, S. (二本柳聡史), Yamaguchi, S. (山口祥一) & Tahara, T. (田原太平). Direct evidence for orientational flip-flop of water molecules at charged interfaces: A heterodyne-detected vibrational sum frequency generation study. *J. Chem. Phys.* **130**, 204704 (2009).

36. Wang, Y., Seki, T., Liu, X., Yu, X., Yu, C. C., Domke, K. F., ... & Bonn, M. Direct Probe of Electrochemical Pseudocapacitive pH Jump at a Graphene Electrode**. *Angew. Chem. Int. Ed.* **62**, e202216604 (2023).

37. Ohno, K., Okimura, M., Akai, N. & Katsumoto, Y. The effect of cooperative hydrogen bonding on the OH stretching-band shift for water clusters studied by matrix-isolation infrared spectroscopy and density functional theory. *Phys. Chem. Chem. Phys.* **7**, 3005–3014 (2005).

38. Yu, C. C., Seki, T., Chiang, K. Y., Tang, F., Sun, S., Bonn, M., & Nagata, Y. Polarization-Dependent Heterodyne-Detected Sum-Frequency Generation Spectroscopy as a Tool to Explore Surface Molecular Orientation and Ångström-Scale Depth Profiling. *J. Phys. Chem. B* **126**, 6113–6124 (2022).

39. Chiang, K. Y., Seki, T., Yu, C. C., Ohto, T., Hunger, J., Bonn, M., & Nagata, Y. The dielectric function profile across the water interface through surface-specific vibrational spectroscopy and simulations. *Proc. Natl Acad. Sci. USA* **119**, e2204156119 (2022).

40. Xu, Y., Ma, Y.-B., Gu, F., Yang, S.-S. & Tian, C.-S. Structure evolution at the gate-tunable suspended graphene–water interface. *Nature* **621**, 506–510 (2023).

41. Wen, Y. C., Zha, S., Liu, X., Yang, S., Guo, P., Shi, G., ... & Tian, C. Unveiling Microscopic Structures of Charged Water Interfaces by Surface-Specific Vibrational Spectroscopy. *Phys. Rev. Lett.* **116**, 016101 (2016).

42. Ohno, P. E., Wang, H. & Geiger, F. M. Second-order spectral lineshapes from charged interfaces. *Nat Commun* **8**, 1032 (2017).

43. Reddy, S. K., Thiraux, R., Rudd, B. A. W., Lin, L., Adel, T., Joutsuka, T., ... & Paesani, F. Bulk Contributions Modulate the Sum-Frequency Generation Spectra of Water on Model Sea-Spray Aerosols. *Chem* **4**, 1629–1644 (2018).

44. Yu, X., Chiang, K.-Y., Yu, C.-C., Bonn, M. & Nagata, Y. On the Fresnel factor correction of sum-frequency generation spectra of interfacial water. *J. Chem. Phys.* **158**, 044701 (2023).

45. Zhuang, X., Miranda, P. B., Kim, D. & Shen, Y. R. Mapping molecular orientation and conformation at interfaces by surface nonlinear optics. *Phys. Rev. B* **59**, 12632–12640 (1999).

46. Odendahl, N. L. & Geissler, P. L. Local Ice-like Structure at the Liquid Water Surface. *J. Am. Chem.*

*Soc.* **144**, 11178–11188 (2022).

47. Li, Q., Song, J., Besenbacher, F. & Dong, M. Two-Dimensional Material Confined Water. *Acc. Chem. Res.* **48**, 119–127 (2015).

48. Yang, S., Zhao, X., Lu, Y. H., Barnard, E. S., Yang, P., Baskin, A., ... & Salmeron, M. Nature of the Electrical Double Layer on Suspended Graphene Electrodes. *J. Am. Chem. Soc.* **144**, 13327–13333 (2022).

49. Wang, Y., Tang, F., Yu, X., Ohto, T., Nagata, Y., & Bonn, M. Heterodyne-Detected Sum-Frequency Generation Vibrational Spectroscopy Reveals Aqueous Molecular Structure at the Suspended Graphene/Water Interface. *Angew. Chem. Int. Ed.* **n/a**, e202319503 (2024).

50. Kühne, T. D., Iannuzzi, M., Del Ben, M., Rybkin, V. V., Seewald, P., Stein, F., ... & Hutter, J. CP2K: An electronic structure and molecular dynamics software package - Quickstep: Efficient and accurate electronic structure calculations. *J. Chem. Phys.* **152**, 194103 (2020).

51. Wang, H., Zhang, L., Han, J. & E, W. DeePMD-kit: A deep learning package for many-body potential energy representation and molecular dynamics. *Comput. Phys. Commun.* **228**, 178–184 (2018).

52. Thompson, A. P., Aktulga, H. M., Berger, R., Bolintineanu, D. S., Brown, W. M., Crozier, P. S., ... & Plimpton, S. J. LAMMPS - a flexible simulation tool for particle-based materials modeling at the atomic, meso, and continuum scales. *Comput. Phys. Commun.* **271**, 108171 (2022).

53. Ohto, T., Usui, K., Hasegawa, T., Bonn, M. & Nagata, Y. Toward ab initio molecular dynamics modeling for sum-frequency generation spectra; an efficient algorithm based on surface-specific velocity-velocity correlation function. *J. Chem. Phys.* **143**, 124702 (2015).

# Supplementary Information for:

# Interfaces Govern Structure of Angstrom-Scale Confined Water

*Yongkang Wang,[1,2#] Fujie Tang,[3,4#] Xiaoqing Yu,[1] Kuo-Yang Chiang,[1] Chun-Chieh Yu,[1] Tatsuhiko Ohto,[5] Yunfei Chen,[2] Yuki Nagata,[1*] Mischa Bonn[1*]*

[1] *Max Planck Institute for Polymer Research, Ackermannweg 10, 55128 Mainz, Germany.*
[2] *School of Mechanical Engineering, Southeast University, 211189 Nanjing, China.*
[3] *Pen-Tung Sah Institute of Micro-Nano Science and Technology, Xiamen University, 361005 Xiamen, China.*
[4] *Laboratory of AI for Electrochemistry (AI4EC), IKKEM, 361005 Xiamen, China*
[5] *Graduate School of Engineering, Nagoya University, Nagoya 464-8603, Japan.*
[#] *Yongkang Wang and Fujie Tang contributed equally to this work.*

[*] *Correspondence to: nagata@mpip-mainz.mpg.de, bonn@mpip-mainz.mpg.de*

**Supplementary Information contains:**

Supplementary Methods Sections 1-11

Supplementary Discussion Sections S1 to S14

Figs. S1 to S25

Table S1

# Contents

# Supplementary Methods

## 1. Chemicals

All related chemicals of lithium chloride (LiCl), heavy water ($D_2O$), hydrochloride (HCl, 37%), concentrated sulfuric acid ($H_2SO_4$, 98%), 30 wt. % hydrogen peroxide solution ($H_2O_2$), ammonium persulfate ($(NH_4)_2S_2O_8$), cellulose acetate butyrate (CAB), ethyl acetate, 2-propanol, and acetone were purchased from Sigma-Aldrich and were all of analytical grade without further purification. Deionized water was provided by a Milli-Q system (resistivity ≥ 18.2 MΩ·cm and TOC ≤ 4 ppb) and was saturated with argon by bubbling gas through it for 30 minutes before use. CVD-grown graphene on copper foils was purchased from Grolltex Inc.

## 2. Substrate Preparation

$CaF_2$ substrates (25 mm diameter with a thickness of 2 mm, Korth Crystals GmbH) and $SiO_2$ substrates (25 mm diameter with a thickness of 2 mm, PI-KEM Ltd) were cleaned with acetone and 2-propanol sequentially. Ultrasonic cleaning was avoided to guarantee a flat $CaF_2$ surface. Note that the facet of the $CaF_2$ and $SiO_2$ substrates is unclear. After that, 100 nm-thick gold film was thermally evaporated onto the $CaF_2$ and $SiO_2$ substrates with a shadow mask. The gold film serves as the reference sample to generate a stable and precise reference phase for the HD-SFG measurements. Additionally, the gold film also serves as a marker for the AFM and HD-SFG measurements.

## 3. Nanoconfined Water Sample Preparation

The nanoconfined water samples were prepared by using the wet transfer technique to enclose water between a graphene sheet and a $CaF_2$ substrate. Detailed procedures are shown in Fig. S1. After the sample preparation, a thin film of LiCl solution was trapped between the graphene and the substrate due to the hydrophilic nature of the $CaF_2$ substrate. The prepared samples were dried at room temperature for more than 12 hours to remove the excess water. Finally, the CAB layer was removed by immersing the sample in acetone. Before assembling the sample onto the sample cell, the prepared sample was allowed to be stored for ~6 hours at RH ~25% for evaporation of acetone. Using this method, a centimeter-sized nanoconfined water sample was obtained (Fig. 1a). Two-dimensional nanoconfined water between the graphene sheet and SiO$_2$ substrate was prepared using the same procedures.

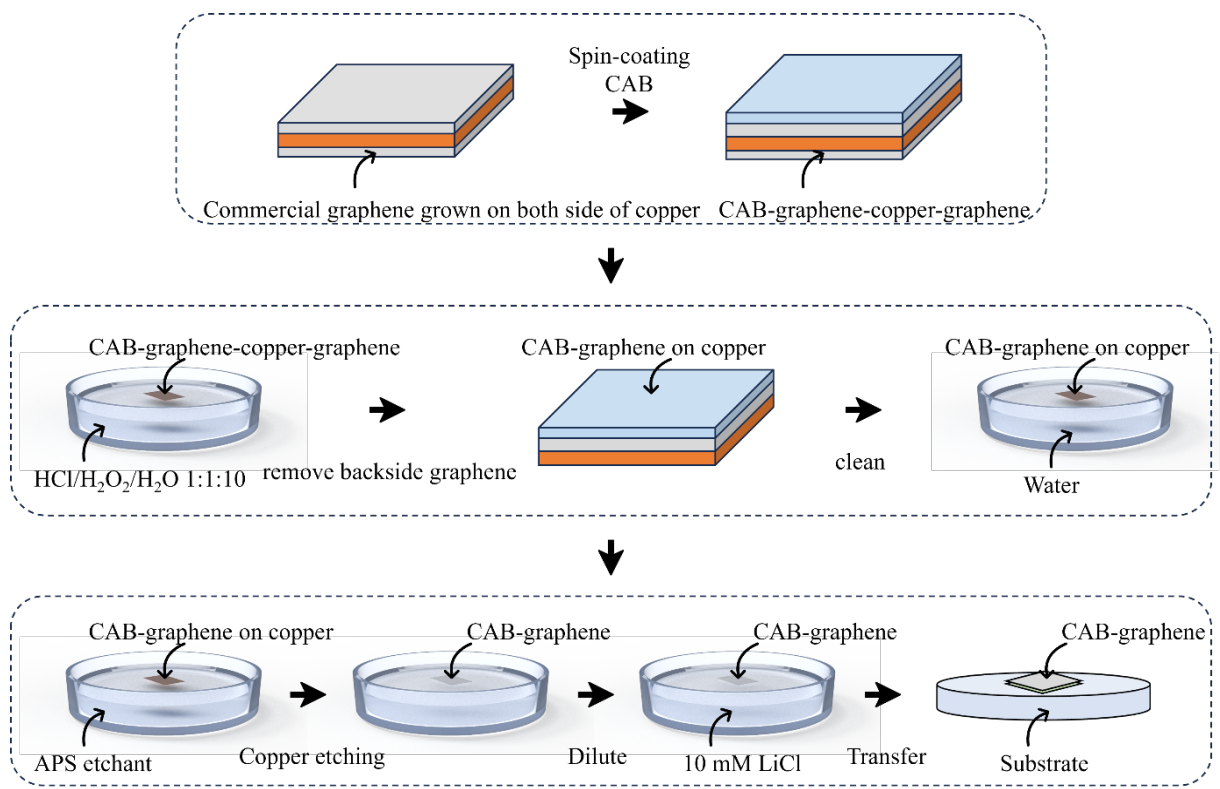

Fig. S1 | **Procedures for the preparation of nanoconfined water sample.**

## 4. Suspended Graphene on the Water Surface

The preparation of the suspended graphene on the water surface was similar to Refs.[1–3] Detailed procedures are summarized in Fig. S2. In brief, the graphene layer grown on the backside of the copper foil was first removed using HCl/H$_2$O$_2$/H$_2$O mixture solution. Then, the CAB-coated graphene on the copper foil was immersed in acetone to remove the CAB layer. Subsequently, the copper foil was exposed to a 10 mM APS solution for more than 12 hours to etch away the copper foil. Then the solution was recycled with pure water several times to dilute and remove the etching agents. Finally, the solution was replaced by the electrolyte (LiCl) solution to reach the target concentrations. An O-ring was used to trap the prepared graphene sample. Using this method, centimeter-sized monolayer graphene suspended on the water surface was obtained (Fig. S2).

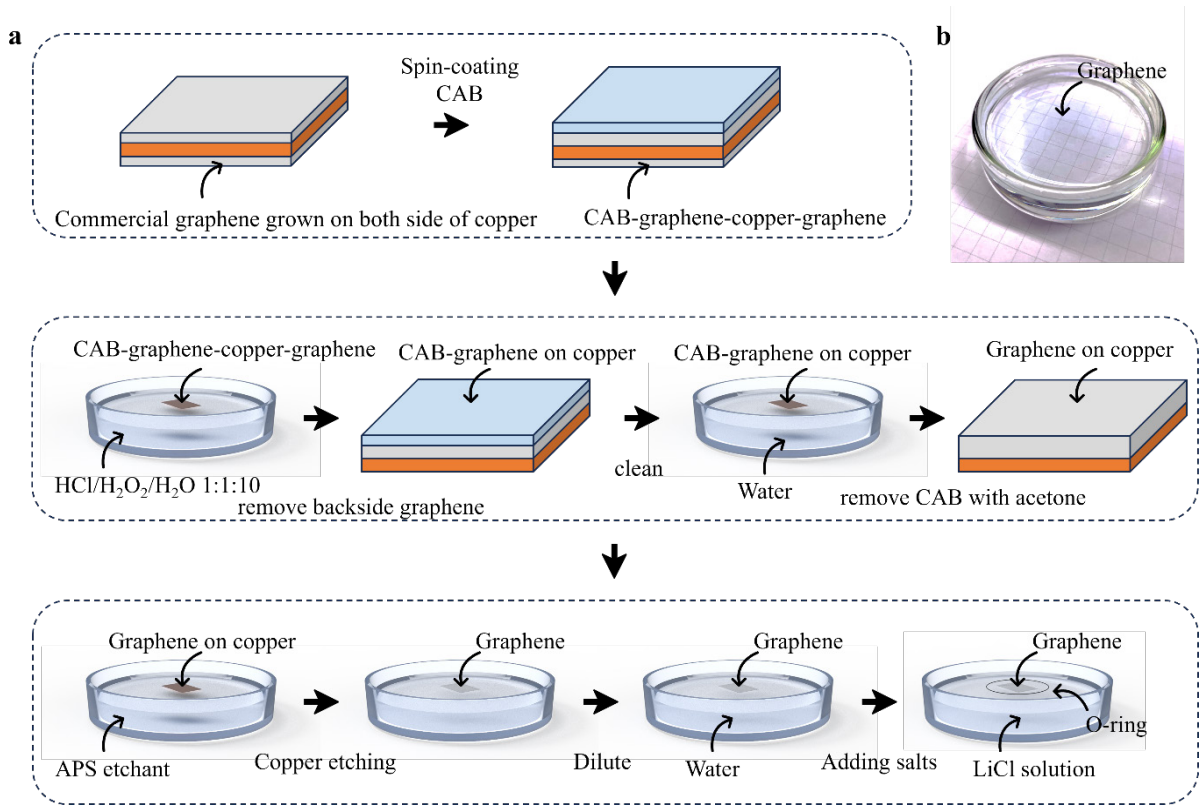

**Fig. S2 | Preparation of suspended graphene on the water surface. a.** Sample preparation procedures. **b.** A photo of the centimeter-sized suspended graphene on the water surface. The diameter of the petri dish is 5 cm.

## 5. Sample Cell

Our sample cell is schematically depicted in Fig. 1f and Fig. S3a. A photo of the cell is shown in Fig. S3b. The cell mainly consists of two rectangular polytetrafluoroethylene (PTFE) parts, the top clamp part, and the bottom flowing channel (~12×3×3 mm$^3$) part. The top clamp has an opening of ~16 mm in diameter for the light beam paths. The bottom part has two round holes on the left- and right-side walls, serving as the inlet and outlet of nitrogen (N$_2$), respectively. Another hole near the cell outlet was connected to a humidity sensor to measure the RH inside the sample cell. The prepared sample and an O-ring were then sandwiched between the top and the bottom PTFE parts. The O-ring creates a seal between the N$_2$ or aqueous solution and the air. The base and clamp parts were cleaned with piranha solution before use.

The RH in the cell was tuned by purging the cell with $N_2$ of different RHs. The RH of the $N_2$ was tuned by changing the relative flow rate of a mixture comprising dry $N_2$ and wet $N_2$ while maintaining the overall $N_2$ flow rate in the sample cell (Fig. S3c). To guarantee the RH control, a microscopic humidity sensor was connected to the sample cell near the cell outlet to measure the RH in the sample cell (Fig. S3b).

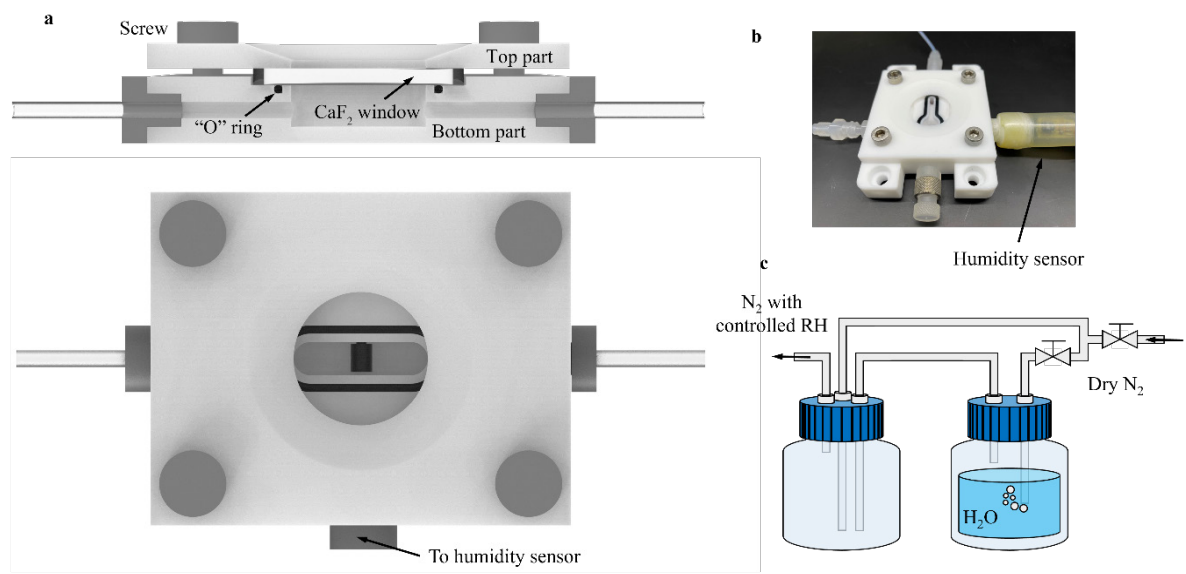

**Fig. S3 | Experimental setup. a.** Schematic diagram of the sample cell for HD-SFG measurements. **b.** A photo of the sample cell. **c.** Tuning RH of $N_2$.

## 6. Raman Measurement

The Raman spectra were recorded with a WITec confocal Raman spectrometer (Alpha 300 R, ×10 objective) with 600 grooves/mm grating, 532 nm laser, 2 mW power, and 10 s integration time.

## 7. HD-SFG Measurement

Our non-collinear HD-SFG setup was built with a Ti: Sapphire regenerative amplifier laser system. A detailed description can be found in Refs.[4,5]. For the nanoconfined water sample and suspended graphene sample, each spectrum was acquired with an exposure time of 2 minutes and measured fifteen times on average with a total exposure time of 30 minutes. The power of the IR and visible beams was reduced to below 3 mW to avoid burning the graphene sheet. For the $CaF_2$/water interface, the power of the IR and visible beams was around 4 mW and 8 mW, respectively. Each spectrum was acquired with an exposure time of 10 minutes and measured 6 times on average with a total exposure time of 60 minutes. The nanoconfined water sample and CaF$_2$/water sample HD-SFG spectra at *ssp* polarization were normalized with that for the CaF$_2$/gold at *ssp* polarization. The suspended graphene sample HD-SFG spectra at *ssp* polarization were normalized with that for the air/*z*-cut quartz (zqz) at *ssp* polarization.

## 8. Calibration of Experimental $\chi^{(2)}$ Spectra

The total intensity of the HD-SFG signal is represented by the sum of sample sum frequency (SF) light and local oscillator (LO) field reflected at an interface, which can be expressed as follows:

$$|E_{\text{total}}|^2 = |E_{\text{sample}}|^2 + |E_{\text{LO}}|^2 + E_{\text{sample}} r^*_{sample} E^*_{\text{LO}} e^{i\omega T} + E^*_{\text{sample}} r_{sample} E_{LO} e^{-i\omega T}. \quad (S1)$$

where $E_{\text{sample}}$ and $E_{\text{LO}}$ are the electric fields of the SF light from the sample and the LO, respectively. $r_{\text{sample}}$ is the reflectivity coefficient. $T$ is the time delay between SF lights from the sample SF and the LO fields. * represents the complex conjugate. The third term on the right-hand side of Eq. S1 was picked up using time-domain filtration (A combination of a boxcar function given by Eq. S2 with $t_c$ the cutoff time and a Happ-Genzel function given by Eq. S3[6]) with Fourier transform and was then converted to an interferogram through inverse Fourier transform.[7] The measured complex-valued spectra of second-order nonlinear susceptibility ($\chi^{(2)}_{ssp,\text{measured}}$) of the nanoconfined water sample were obtained via the Fourier analysis of the obtained interferogram and normalization with that of the CaF$_2$/gold interface (Eq. S4).

$$A(t) = \begin{cases} 1 \text{ for } t \geq t_c \\ 0 \text{ for } t < t_c \end{cases}, \quad (S2)$$

$$A(t) = 0.54 + 0.46 \cos\left(\frac{\pi(t + 0.9 \text{ ps})}{0.5 \text{ ps}}\right), \quad (S3)$$

$$\chi^{(2)}_{ssp,\text{measured}} = \frac{E_{\text{sample}} r^*_{\text{sample}} E^*_{\text{LO}} e^{i\omega T}}{E_{\text{gold}} r^*_{\text{gold}} E^*_{\text{LO}} e^{i\omega T}}, \tag{S4}$$

The signal of the CaF$_2$/gold interface was collected at the gold region of the sample immediately before the sample measurement to ensure a precise reference phase. Additionally, the phase of the gold film was determined by measuring the O-H stretching Im$\chi^{(2)}_{ssp}$ spectrum of D$_2$O at the interface via normalization of the signal with that of the CaF$_2$/gold sample. As D$_2$O does not have any vibrational response in the O-H stretching region,[8] we can determine the phase of the gold surface by the fact that the Im$\chi^{(2)}_{ssp}$ spectrum shows a flat zero line.[5] The HD-SFG spectra at the air/graphene/water interface and the air/D$_2$O interface were obtained via normalization of the signal with that of the air/zqz interface.

### 9. *Ab Initio* Molecular Dynamics (AIMD) Simulation

We carried out SFG spectra simulations for the CaF$_2$(111)/water interface, the water/graphene interface, and the nanoconfined water systems using *ab initio* molecular dynamics (AIMD) simulations with the mixed Gaussian and plane wave approach as implemented in the CP2K code.[9,10] We used the (111) facet as the model of the CaF$_2$ surface because this facet provides the thermodynamically most stable surface.[11] We used the revPBE[12,13] exchange-correlation (XC) functionals together with the empirical van der Waals (vdW) correction scheme of Grimme's D3(0)[14] method. The core electrons were described using the Norm-conserving Goedecker-Teter-Hutter pseudopotentials.[15,16] We employed the short-ranged MOLOPT double-valance $\zeta$ basis with one set of polarization functions (DZVP) and the plane wave density cutoff of 400 Ry. The time step for integrating the equation of motion was set to 0.5 fs. All simulations were performed at 300 K in the NVT ensemble with the thermostat of the canonical sampling through the velocity rescaling method.[17] The surface dipole correction was used to remove the impact of the macroscopic dipole in the image cell.[18]

We ran AIMD simulations for the three nanoconfined systems with different fractions of water, the CaF$_2$/aqueous LiCl solution, and the aqueous LiCl solution/graphene interfaces (Fig. S4). For all the systems, we used the same simulation cell, where the cell vectors were $\vec{a} =$

$(14.76 \text{ Å}, 0 \text{ Å}, 0 \text{ Å})$, $\vec{b} = (7.38 \text{ Å}, 12.78 \text{ Å}, 0 \text{ Å})$, and $\vec{c} = (0 \text{ Å}, 0 \text{ Å}, 50 \text{ Å})$. $\vec{a}$ and $\vec{b}$ formed the surface, while $\vec{c}$ formed the surface normal. This surface unit cell ($\vec{a}$ and $\vec{b}$) corresponds to the graphene-$p$(6×6) hexagonal cell. The lattice mismatch between the graphene sheet and the CaF$_2$(111) surface was compensated by compressing the CaF$_2$(111)-$p$(4×4) by 4.5%. The CaF$_2$ consisted of three layers (one layer is composed of 16 formula units of CaF$_2$), while the graphene sheet was composed of 72 carbon atoms. Since the CaF$_2$(111) surface is positively charged, we detached one fluorine atom from the topmost surface of CaF$_2$ in contact with water and made the fluoride solvated in water to maintain the charge neutrality of the total system. This resulted in a surface charge density of ~98 mC/m$^2$, which is much larger than the experimentally estimated surface charge (40±10 mC/m$^2$, see Supplementary Discussion S1 for more details). This was overcome by utilizing the machine learning technique (see Supplementary Method Section 10).

We prepared the three samples for the nanoconfined water systems. The sample containing the smallest fraction of water (the thickness of the water layer: ~8 Å) was composed of 49 H$_2$O molecules and 2 pairs of Li$^+$ and Cl$^-$ ions, the sample containing the second smallest fraction of water (the thickness of water layer: ~12 Å) was composed of 77 H$_2$O molecules and 3 pairs of Li$^+$ and Cl$^-$ ions, and the sample containing the largest fraction of water (the thickness of water layer: ~16 Å) was composed of 99 H$_2$O molecules and 4 pairs of Li$^+$ and Cl$^-$ ions. The resulting LiCl concentrations for the three nanoconfined water systems were around 2 M. For the CaF$_2$/water sample, we contained 159 H$_2$O molecules and 3 pairs of Li$^+$ and Cl$^-$ ions. The resulting LiCl concentration was ~1 M. For the water/graphene sample, we contained 136 H$_2$O molecules without adding ions because the water spectrum at the water/graphene interface is insensitive to ions. Furthermore, we carried out the simulation without Li$^+$ and Cl$^-$ ions at the CaF$_2$/water interface. Note that in this simulation, a fluoride is solvated in the water, to achieve the charge neutrality of the system. For all the systems, the resulting thickness of the vacuum regions in the simulation box was >15 Å (Fig. S4). As a result, the other side of the interfaces for the CaF$_2$/water and water/graphene samples, which are generated in the slab geometry is the water/air interface. We generated 5 independent configurations for each sample. We equilibrated the systems by running 5 ps, and we sampled the >30 ps trajectories, from which we computed the SFG spectra and performed the analysis.

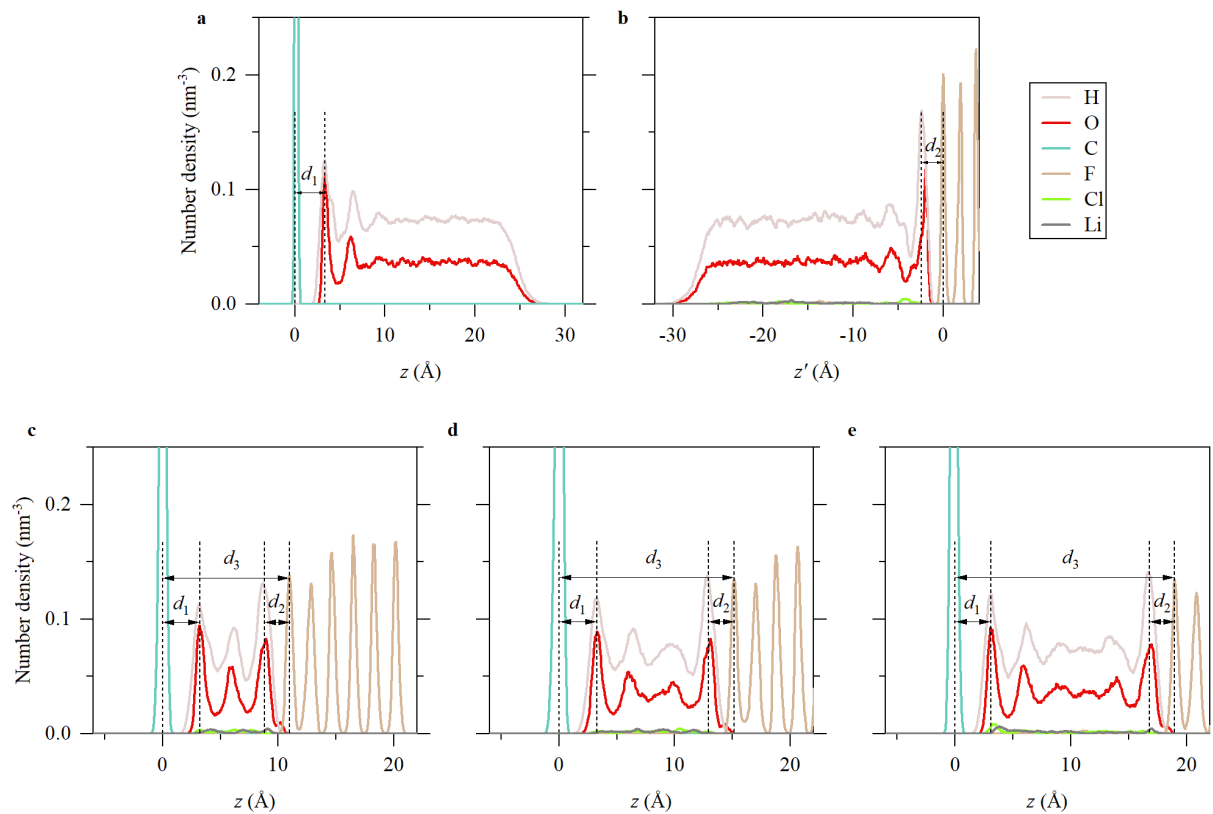

**Fig. S4 | The axial profiles of the number density along the surface normal axis for (a) the water/graphene interface, (b) the $CaF_2$/water interface system as well as (c-e) the nanoconfined systems with a water thickness of (c) ~8 Å, (d) ~12 Å, and (e) ~16 Å.** The origin points of the z-axis and z'-axis were set to the position of the graphene sheet (**a**, **c**, **d**, **e**) and the position of the first layer of F atom (**b**) in the cell, respectively. The obtained distances between the topmost water layers and the graphene sheet are (**a**) $d_1 = 3.3 \pm 0.1$ Å, (**c**) $d_1 = 3.2 \pm 0.1$ Å, (**d**) $d_1 = 3.3 \pm 0.1$ Å, (**e**) $d_1 = 3.2 \pm 0.1$ Å. The distances between the water slab and $CaF_2$ are (**b**) $d_2 = 2.3 \pm 0.1$ Å, (**c**) $d_2 = 2.3 \pm 0.1$ Å, (**d**) $d_2 = 2.3 \pm 0.1$ Å, (**e**) $d_2 = 2.3 \pm 0.1$ Å. The distances between the graphene and $CaF_2$ in the confined water systems are (**c**) $d_3 = 11.0 \pm 0.1$ Å, (**d**) $d_3 = 15.2 \pm 0.1$ Å, (**e**) $d_3 = 19.2 \pm 0.1$ Å.

## 10. Machine Learning Force Field MD Simulation

The surface charge of the $CaF_2$/water interface for the $CaF_2$ substrate employed in our SFG measurements was experimentally estimated to be $40\pm10$ mC/m$^2$ at a neutral pH condition (see

Supplementary Discussion S1). This value is smaller than the surface charge density used in the AIMD simulation; to simulate the $CaF_2$/water interface and the nanoconfined water system with the experimentally estimated surface charge density, one needs to expand the simulation cell size. However, due to the huge computational cost of the AIMD simulation, it is highly challenging to scale up the system size. To overcome the difficulty, we used the machine learning force field (MLFF) based molecular dynamics simulation technique.

We constructed a deep potential model to fit the potential energy surface (PES) which is implemented in the DeePMD-kit package[19–21] for the $CaF_2$/water system as well as the nanoconfined water system. We used the "se_e2_a" descriptor model with local environment cutoff and smooth cutoff parameters set as 6.0 and 0.5 Å. The embedding bet was composed of the hidden layers with their sizes of (25, 50, 100), and the fitting net had the hidden layers with their size of (240, 240, 240). We constructed 4 MLFF models using a concurrent learning workflow implemented in the DP-GEN package.[22] We randomly selected 2800 structures from the AIMD simulated data to train 4 initial models. We repeated the "exploration", "labelling", and "training" process for the large cells explained below until the deviations among the MLFF models were below 0.2 eV/Å for 99% of structures in the sampling trajectories. After that, a final model was trained based on all the training data. The resulting FF model vs *ab initio* data are shown in Fig. S5. Note that, we did not employ the DeePMD with long-range corrections to fit the PES, because the calculated spectra without the long-range corrections show a good agreement with experimental data with current ions concentration. (see Supplementary Discussion S13).

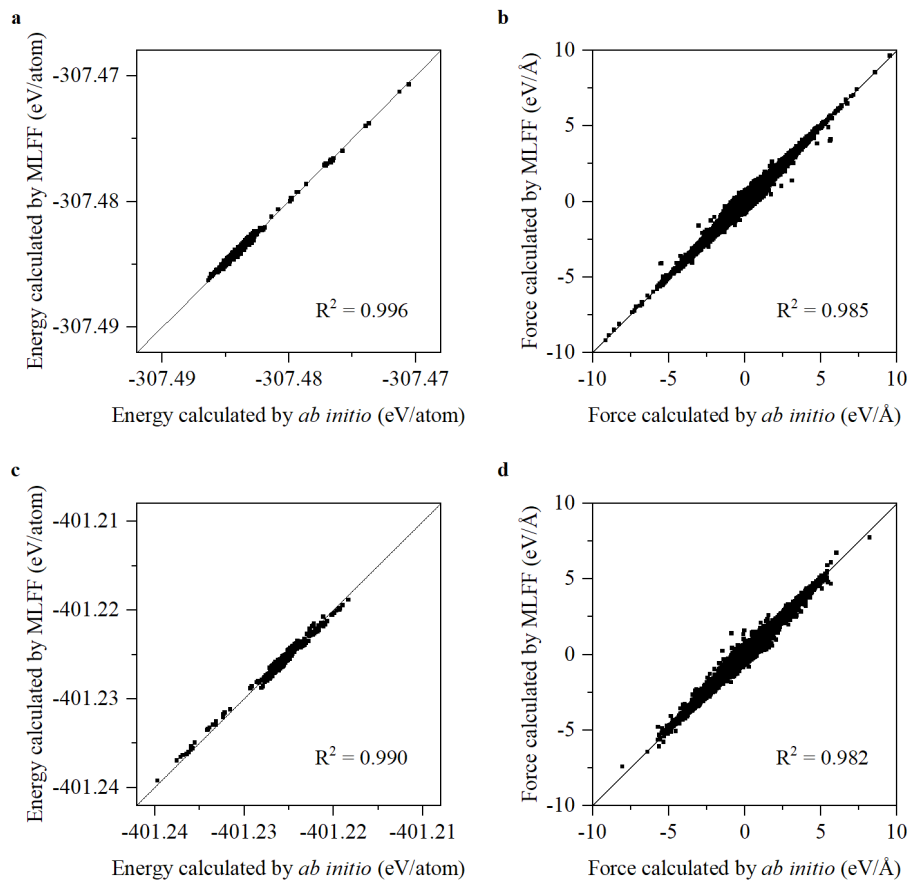

**Fig. S5 | Comparisons of energies and forces obtained by *ab initio* method and MLFF on the test data set of CaF$_2$/water (a-b) and nanoconfined water system (c-d).** The coefficient of determination ($R^2$) is shown in each plot.

We constructed the cell composed of $2\vec{a}$, $2\vec{b}$, and $\vec{c}$, where we detached two fluorine atoms from the CaF$_2$ surface and made the fluorine ions solvated in water. This led to a surface charge density of 49 mC/m$^2$. This surface charge value is consistent with the experimentally obtained value (40±10 mC/m$^2$). For the CaF$_2$/water sample, we contained 596 H$_2$O molecules and 12 pairs of Li$^+$ and Cl$^-$ ions. For the nanoconfined water sample, we contained 196 H$_2$O molecules and 8 pairs of Li$^+$ and Cl$^-$ ions. The nanoconfined water system provides a water layer thickness of ~8.0 Å, close to the AFM data of ~8.2 Å. We also constructed another nanoconfined water sample using a higher ion concentration with the same cell size, which contained 172 H$_2$O molecules and 20 pairs of Li$^+$ and Cl$^-$ ions. This combination increases the LiCl concentration thrice, from ~2 M to ~6 M.

After generating the MLFF model, we carried out the force field MD simulation for $CaF_2$/water and nanoconfined water systems of high and low LiCl concentrations by using the LAMMPS package.[23] The equation of motion was integrated with a timestep of 0.5 fs. We used the Nose-Hoover thermostat with a temperature damping parameter of 1 ps. We ran the MLFF-MD simulation with NVT ensemble for the total 5 independent systems at 300 K for low LiCl concentration with total trajectories length of >2 ns. and a total of four independent systems at 300 K for high concentration with total trajectories length of 4 ns. From all the MD trajectories, we calculated the SFG spectra and performed the analysis.

## 11. $\chi^{(2)}$ Spectrum Calculation

We computed the SFG spectra by using the AIMD or MLFF MD trajectories by using the surface-specific velocity-velocity auto-correlation function (ssVVAF) approach.[24] This method enables one to compute the SFG spectra with a reasonable s/n ratio solely from the MD trajectories. Within this ssVVAF formalism, the resonant part of the SFG susceptibility, $\chi_{xxz}^{(2),R}(\omega)$, can be given as:

$$\chi_{xxz}^{(2),R}(\omega) = \frac{Q(\omega)\mu'(\omega)\alpha'(\omega)}{i\omega^2}\chi_{xxz}^{ssVVAF}(\omega), \tag{S5}$$

$$\chi_{xxz}^{ssVVAF}(\omega) = \int_0^\infty dt\, e^{-i\omega t} \langle \sum_i g_{ds}(z_i(0))\dot{r}_{z,i}^{OH}(0)\frac{\dot{\vec{r}}_i^{OH}(t)\cdot\vec{r}_i^{OH}(t)}{|\vec{r}_i^{OH}(t)|}\rangle, \tag{S6}$$

where $g_{ds}(z_i)$ is the truncation function for the dividing surface to selectively extract the vibrational responses of water molecules near the interface:

$$g_{ds}(z_i) = \begin{cases} 0 \text{ for } z_i \geq z_{ds} \\ 1 \text{ for } z_i < z_{ds} \end{cases}, \tag{S7}$$

where $z_{ds}$ is the z-coordinate of the dividing surface and $z_i$ is the z-coordinate of the O atom of the ith O-H bond. The $z_{ds}$ value was set to decouple the responses of the $CaF_2$/water interface and graphene/water interface from the water/air interface. Note that for the confined water system, we did not use any dividing surface; the SFG response was calculated using the whole water slab. We set the origin point as the averaged position of the first layer of F atoms for the CaF$_2$/water interface, and the averaged position of C atoms for the graphene/water interface. The $z_{ds}$ value is set to 12 Å for both the CaF$_2$/water system and the graphene/water system. Due to the geometry, the $g_{ds}(z_i)$ function for CaF$_2$/water system is slightly modified as:

$$g_{ds}(z_i) = \begin{cases} 0 \text{ for } z_i \leq -z_{ds} \\ 1 \text{ for } z_i > -z_{ds} \end{cases}. \tag{S8}$$

The frequency-dependent induced transition dipole moment and polarizability due to the solvation effects were included by using the frequency-dependent transition dipole moment ($\mu'(\omega)$) and polarizability ($\alpha'(\omega)$):[25,26]

$$\mu'(\omega) \equiv \left(1.377 + \frac{53.03(3737.0 - \omega)}{6932.2}\right)\mu^0, \tag{S9}$$

$$\alpha'(\omega) \equiv \left(1.271 + \frac{5.287(3737.0 - \omega)}{6932.2}\right)\alpha^0, \tag{S10}$$

where $\omega$ is in cm$^{-1}$. $\mu^0$ and $\alpha^0$ are permanent dipole moments and permanent polarizability of OH chromophores, respectively. $Q(\omega)$ is the quantum correction factor given by:[27]

$$Q(\omega) = \frac{\beta \hbar \omega}{1 - \exp(-\beta \hbar \omega)}, \tag{S11}$$

where $\beta = 1/kT$ is the inverse temperature.

Because the CaF$_2$/water systems contained both the CaF$_2$/water interface and the water/air interface, we needed to exclude the contribution from the water/air interface. To this end, we computed the layer-by-layer SFG contribution by including the interfacial water layers near the CaF$_2$/water interface by using the AIMD data. The results shown in Fig. S6 indicate that the $z_{ds}$ value is set to 12 Å for the CaF$_2$/water interface is sufficient to exclude the contribution from the water/air interface.

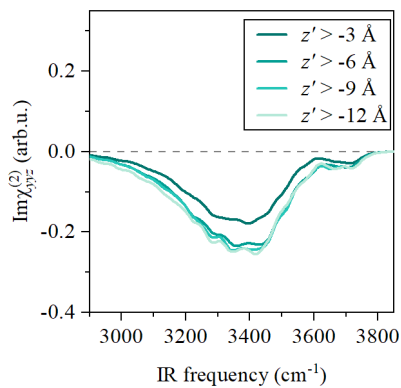

**Fig. S6 | Layer-by-layer contribution of SFG spectra at the CaF$_2$/water interface obtained by using AIMD data.**

This notion was further supported by the dipole orientation of the water molecules (Fig. S7). Even without any Li$^+$ and Cl$^-$ in water and only with an F$^-$, the orientation is saturated at $z$>-12 Å. These data guaranteed that the inclusion of the water molecule in the $z$>-12 Å region is sufficient to reproduce the SFG signal.

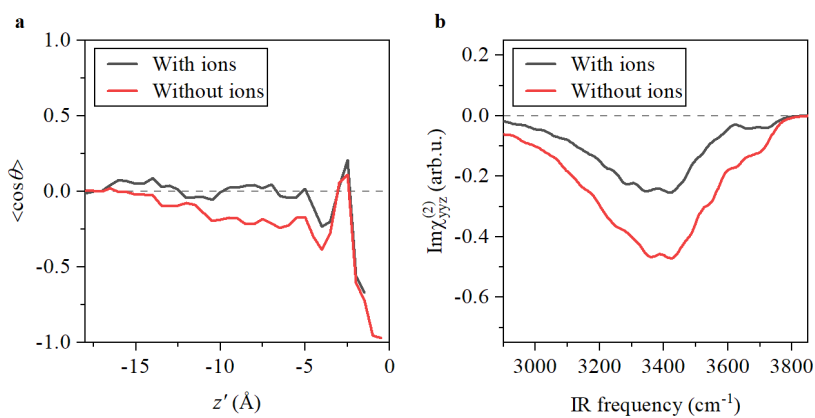

**Fig. S7 | Comparison of (a) dipole orientation and (b) AIMD based $\mathrm{Im}\chi^{(2)}_{yyz}$ spectra of CaF$_2$/water interfaces with and without Li$^+$ and Cl$^-$ ions.**

# Supplementary Discussion

## S1. LiCl Concentration Dependence of SFG Spectra

At a charged interface, an electric field penetrates into the bulk solution and induces alignment and polarization of water molecules in the diffuse layer, providing a bulk contribution to the SFG signal.[28–30] Solvated ion screens the electric field generated at the charged surface, whose concentration, therefore, significantly affects the SFG signal. Such ion concentration dependence is vanishingly weak at high ion concentrations (> 1 M, for example), according to the Gouy-Chapman theory.[30–32] Nevertheless, recent studies proposed that ions of high concentrations could also alter the water structure near the charged surface, affecting the SFG spectra, such as at the $SiO_2$ surface.[32] Here, we examined the sensitivity of the SFG spectra to the LiCl concentration for the $CaF_2$/water interface as well as the nanoconfined water system.

First, we measured $\text{Im}\chi^{(2)}_{yyz,\text{CaF}_2/\text{W}}$ spectra at the $CaF_2$/water interface by varying the LiCl concentrations. The data are displayed in Fig. S8a. Only a minor reduction of the SFG amplitude was observed when increasing the LiCl concentration from 1 M to 8 M at the $CaF_2$/water interface. This can be fully explained by the screening of the surface charge of the $CaF_2$ upon increasing the ion concentration, which is effectively completed when the electrolyte concentration reaches 1 M. Indeed, after removing the bulk contribution ($\chi^{(3)}$) by considering a constant surface charge density ($\sigma_0$) within the Gouy-Chapman theory (Eq. S12),[28–30] the surface contribution ($\chi^{(2)}_{yyz,s}$) exhibits negligible change in both peak intensity and lineshape, confirming that $\sigma_0$ is constant (Fig. S8b). This indicates that the ions do not (or negligibly) induce further surface charging (*i.e.*, change of the density of surface charges) of the $CaF_2$ surface. The negligible surface charging of $CaF_2$ allows us to directly verify the validity of Eq. (1) in our main text and is one of the main reasons we chose the $CaF_2$ substrate to construct the two-dimensional nanoconfined water system.

$$\chi^{(2)}(\sigma_0, c) = \chi^{(2)}_s + \chi^{(3)}\phi_0(\sigma_0, c)\kappa(c)/(\kappa(c) - i\Delta k_z), \tag{S12}$$

where $\chi^{(3)}$ represents the third-order nonlinear susceptibility originating from bulk water, $\phi_0$ is the electrostatic potential, $\kappa$ the inverse of Debye screening length, $c$ electrolyte concentration, and $\Delta k_z$ the phase-mismatch of the SF, visible, and IR beams in the depth direction.

Unlike the CaF$_2$ surface, SFG signals at the SiO$_2$ surface are sensitive to the ion concentration, probably because of the ion-induced further surface charging.[32] To confirm this, we measured $\text{Im}\chi^{(2)}_{yyz,\text{SiO}_2/\text{W}}$ spectra at the SiO$_2$/water interface by varying the LiCl concentrations. The data are displayed in Fig. S8c. Consistent with Ref.[32], the $\text{Im}\chi^{(2)}_{yyz,\text{SiO}_2/\text{W}}$ spectra vary significantly with the high ion concentration, implying that ion-induced surface charging occurs at the SiO$_2$/water interface. Remarkably, surface charging of the SiO$_2$ allows us to estimate the ion concentration of the confined LiCl solution by comparing the confined water signal with the sum signal of signals at the bulk water/graphene interface and the SiO$_2$/bulk water interface at different LiCl concentrations. The data is shown in Fig. S9. The SFG signal of water confined between graphene and SiO$_2$ substrate matches the sum signal of the water/graphene interface and SiO$_2$/water interface signals at a LiCl concentration of ~2-3 M. Assuming that the SiO$_2$ undergoes the same ion-induced charging behavior in nanoconfined water system and SiO$_2$/bulk water system, we conclude that the LiCl concentration of the nanoconfined water in such two-dimensional nanoconfined water system is ~2-3 M.

We further examined the LiCl concentration dependence of the SFG spectra for the nanoconfined system. Here, we employed the machine learning MD simulations to examine this, because our sample preparation method could control the LiCl concentration only passively. The calculated SFG spectra with 2 M and 6 M LiCl are shown in Fig. S8d. We used the nanoconfined system with a thickness of 8 Å. The SFG spectra are insensitive to ion concentration even in the nanoconfined system.

In contact with water (at neutral pH), the CaF$_2$ surface is positively charged.[5,33] Here, we estimated the surface charge density ($\sigma_0$) of the CaF$_2$ substrate. To this end, we measured the SFG signals at the CaF$_2$/water interface at different ion concentrations ($c_1 = 10$ mM and $c_2 =$ 2 M) and then estimated the $\sigma_0$ from the differential spectrum $\Delta\chi^{(2)} = \chi^{(2)}(\sigma_0, c_1) - \chi^{(2)}(\sigma_0, c_2)$ within the Gouy-Chapman theory.[5,28–30] The $\text{Im}\Delta\chi^{(2)}$ spectrum is shown in Fig. S10 and corresponding experimentally estimated $\sigma_0$ is 40±10 mC/m².

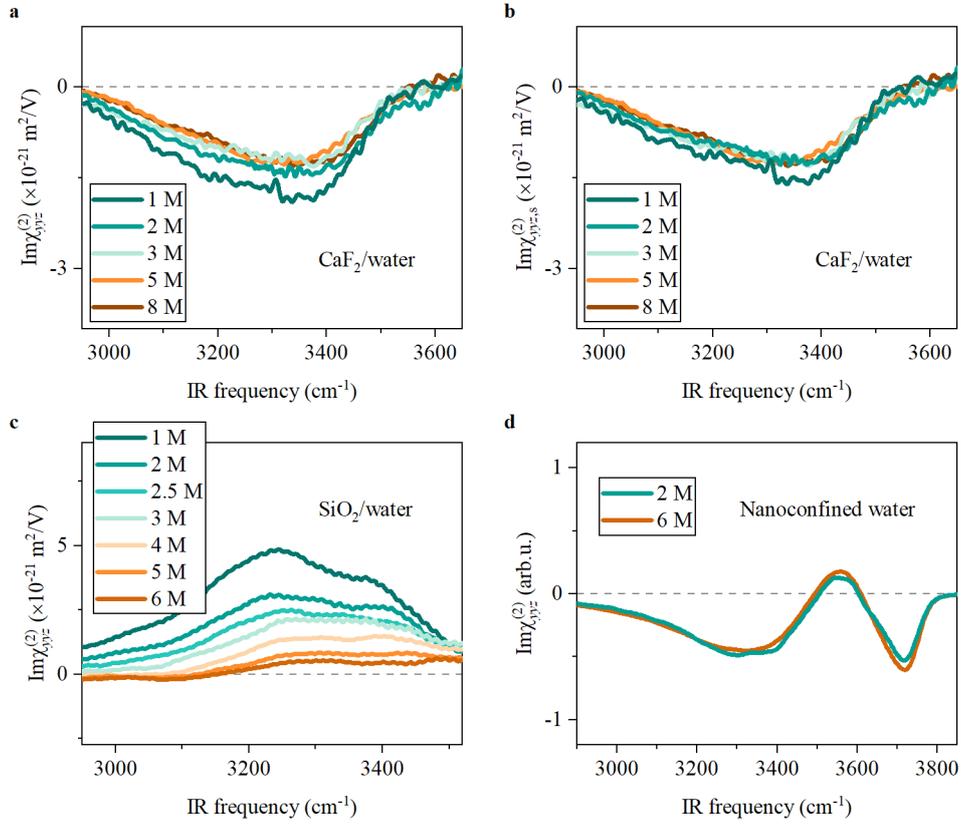

**Fig. S8 | Effect of LiCl concentration. a, c**. The $\text{Im}\chi^{(2)}_{yyz}$ spectrum at the (**a**) CaF$_2$/bulk water interface and (**c**) SiO$_2$/bulk water interface at different LiCl concentrations. **b**. The $\text{Im}\chi^{(2)}_{yyz,s}$ spectrum at the CaF$_2$/bulk water interface at different LiCl concentrations. **d**. Simulated SFG spectra with different ion concentrations for the ~8 Å nanoconfined system computed from MLFF-MD data. The dashed lines in (**a-d**) serve as the zero lines.

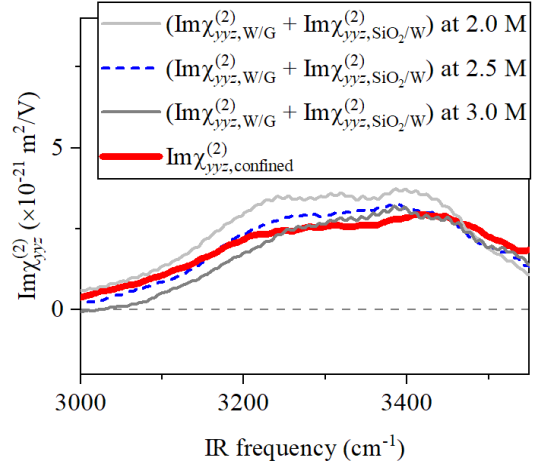

**Fig. S9 | Estimation of ion concentration under confinement.** Experimental $\mathrm{Im}\chi^{(2)}_{yyz}$ spectra of the nanoconfined water between graphene and $SiO_2$. The sum of the water/graphene and $SiO_2$/water signals obtained using different concentrations of LiCl is also shown for comparison. The grey dashed line serves as a zero line.

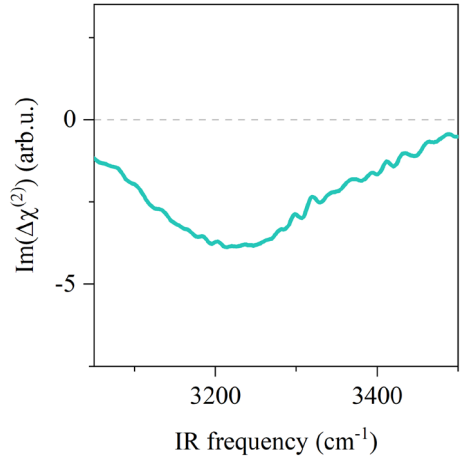

**Fig. S10 | $\mathrm{Im}\Delta\chi^{(2)}$ spectrum for estimation of $\sigma_0$ of the $CaF_2$ substrate.** Corresponding $\sigma_0$ were estimated following the recipe reported in Ref.[5]. The dashed line serves as a zero line.

## S2. Characterization of the CaF$_2$ Surface

To confirm the CaF$_2$ surface is flat and uniform within the HD-SFG probe region (the diameter of the laser spot is around 100 μm), we conducted AFM measurements on the bare CaF$_2$ substrate across a randomly chosen 100×100 μm$^2$ region. The AFM data shows that while some atomic terraces or defects are present (diagonal lines), the CaF$_2$ surface appears atomically flat within the SFG probed region, showing an RMS surface roughness (R$_q$) measuring around 1.3 Å (Fig. S11a). The atomically flat CaF$_2$ surface is further confirmed by the close-up AFM image (0.5×0.5 μm$^2$) showing a R$_q$~1.0 Å comparable to that of a graphite flake (R$_q$~0.8 Å, Figs. S11b, and c). The graphite flake is a good reference since it is a commonly used atomically flat substrate in nanofluidic devices.[34,35] Importantly, the surface roughness is appreciably smaller than a monolayer of water (~3.7 Å).[36] We emphasize that, despite using organic solvents for cleaning the surface, the CaF$_2$ surface is clean as seen from the absence of C-H stretch peaks (2850-2950 cm$^{-1}$) from possible organic contamination in the SFG spectrum of the nanoconfined water sample (Fig. S11d).

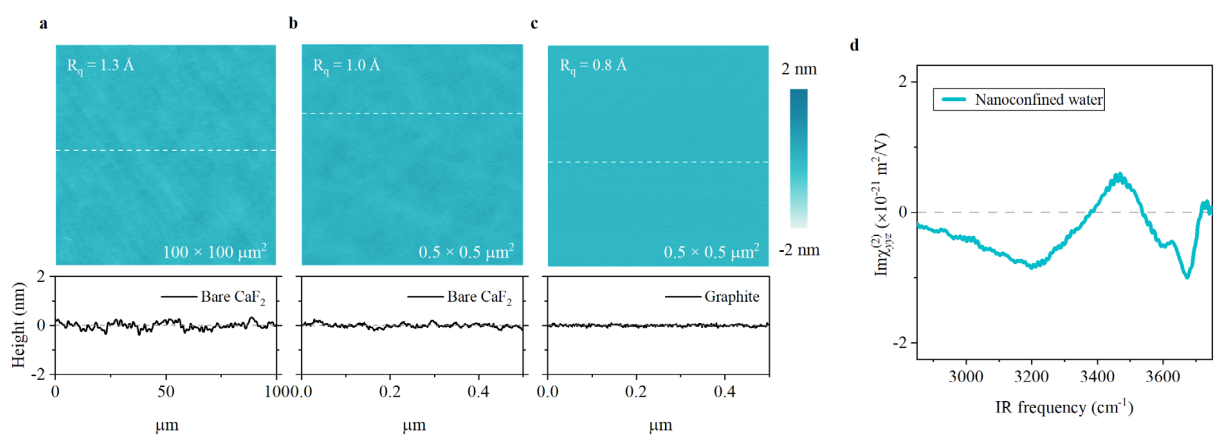

**Fig. S11 | Characterization of the CaF$_2$ surface. a**. AFM height images of the bare CaF$_2$ substrate across a 100×100 μm$^2$ region. **b**, **c**. High-resolution height images of the atomically flat CaF$_2$ substrate and graphite. The bottom panel in each AFM height image shows the typical height profiles along the white dashed lines in the corresponding AFM height image. The dashed grey lines in the height profiles indicate zero lines. R$_q$ values were calculated across the whole scan area. **d**. Cleanliness of the CaF$_2$ substrate. The Im$\chi^{(2)}_{yyz}$ spectrum of the nanoconfined water sandwiched between the graphene sheet and the atomically flat CaF$_2$ substrate. The dashed line serves as a zero line.

## S3. AFM Characterization of the Sample Height

To ensure SFG and AFM probed the same position of the sample, we marked a region of approximately 5000×200 μm² at the edge of the sample by covering the unprobed region with gold. An optical image of the sample is presented in Fig. S12a, while Fig. S12b provides a close-up optical image. AFM and SFG measurements were conducted within this marked sample region (with a sample size of ~250×200 μm² as seen in Fig. S12b). An optical microscope can identify the nanoconfined water region probed with AFM. In the HD-SFG measurement, we identified the marked region by sliding the sample with a translation stage (~10 μm translation accuracy) and checking the HD-SFG signals; in the gold region, the SFG signal from gold is very strong, and in the bare CaF$_2$ region, there is negligible water signal. As such, we could identify the nanoconfined water region by minimizing the signal along the *y*-axis and maximizing the water signal along the *x*-axis, where *x*- and *y*-axes are defined in Fig. S12b. The AFM data measured from the marked region is shown in Fig. S12c. Across the 100×100 μm² scan region, although some graphene wrinkles appear, the nanoconfined water spreads flat and uniform on the CaF$_2$ substrate. The height of the graphene together with the nanoconfined water is ~11.5 Å (Fig. S12d). Given that the thickness of the exclusive volume of the graphene sheet is 3.3 Å, the thickness of the nanoconfined water ($h$) is ~8.2 Å.

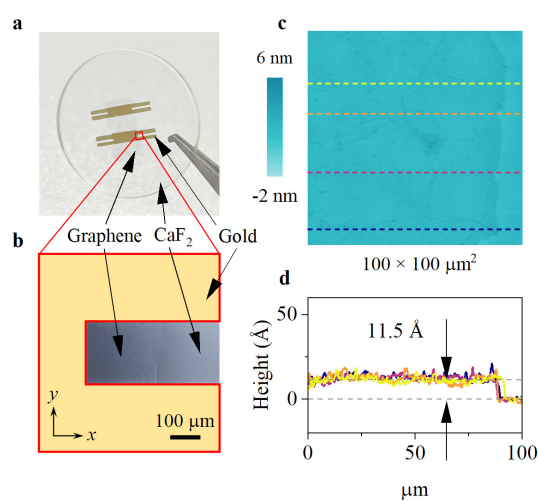

**Fig. S12 | AFM characterization of the nanoconfined water. a**. An optical image of the nanoconfined water sample with a gold marker. **b**. A close-up optical image of the nanoconfined water sample surrounded by gold. **c**. AFM height image of the nanoconfined water sample. **d**. The height profiles along the corresponding-colored dotted lines in (**c**).

## S4. Raman Characterization of the Graphene Sheet

Confinement of water molecules between the graphene and the $CaF_2$ substrate is supposed to induce changes in the strain ($\varepsilon$) and charge density ($n$) of the graphene. The changes in $\varepsilon$ and $n$ will cause the frequency shift of the Raman G-band ($\omega_G$, ~1582 cm$^{-1}$) and 2D-band ($\omega_{2D}$, ~2678 cm$^{-1}$).[37] Accordingly, to probe the effect of confined water on the graphene sheet, we measured the Raman G-band and 2D-band. A typical Raman data is shown in Fig. S13a and the G-band and 2D-band data collected at a randomly selected region are shown in Fig. S13b. A slight blue-shift of $\omega_G$ and $\omega_{2D}$ is observed compared to the graphene supported on a $CaF_2$ substrate without water molecules being confined.

As increasing $n$ ($\varepsilon$) of an intrinsic graphene, its values of ($\omega_G$, $\omega_{2D}$) will move from $O$ along $v_H$ ($v_S$) as shown in Fig. S13b, where $v_H$ ($v_S$) represents the unit vector for $n$ ($\varepsilon$) in the $\omega_G$-$\omega_{2D}$ vector space. $\varepsilon$ and $n$ on the graphene can be independently determined[38,39] through correlation analysis of the frequency shift of $\omega_G$ and $\omega_{2D}$. The variations in $\omega_G$ and $\omega_{2D}$ of the CVD graphene on the $CaF_2$ substrate are purely induced by strain with negligible charge doping ($n < 10^{12}$ cm$^{-2}$). Most graphene experiences a compressive strain varying from 0 to 0.05% due to the presence of wrinkles in CVD graphene. The situation becomes different when water molecules are confined between the graphene monolayer and the $CaF_2$ substrate. In this case, in most regions, graphene experiences a tensile strain ranging from 0 to 0.05% owing to the successful confinement of water molecules. Such a tensile strain enables us to approximately determine the vdW pressure experienced by the trapped solution to be 0 to 500 MPa.[40] Besides the variation along the $v_s$, most data points move towards $v_h$, indicating a slight hole doping ($n^{hole}$ ~ $10^{12}$ cm$^{-2}$). The Raman spectral analysis of the graphene also supports the successful confinement of water molecules between the graphene and the $CaF_2$ substrate.

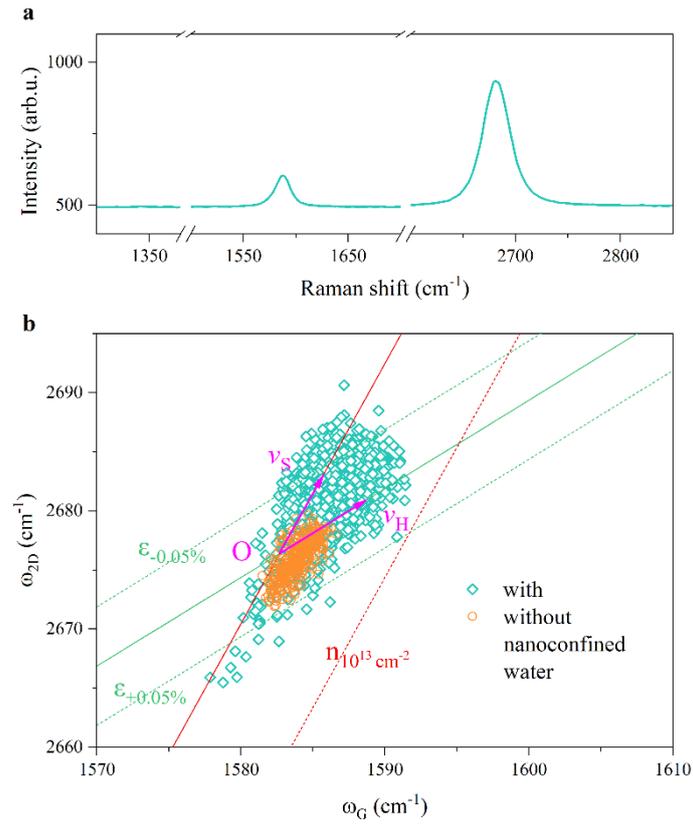

**Fig. S13 | Raman characterization of the graphene sample. a**. Raman spectrum of the graphene layer at a randomly selected region. **b**. Correlation analysis of the frequency shift of $\omega_G$ and $\omega_{2D}$. The red and green lines represent the correlation induced by pure strain and hole doping effects, respectively. Point *O* represents the intrinsic frequencies of the G- and 2D-bands that are not affected by strain or charges and are 1,581.8 cm$^{-1}$ and 2,676.9 cm$^{-1}$, respectively.[37] The averaged strain-sensitivity factor G-band is − 69.1 ± 3.4 cm$^{−1}$/%.[41] The positive charge-sensitivity factor of the G-band is adopted from a reference (red dashed line).[37]

## S5. Capillary Condensation

Our two-dimensional nanoconfined water system, utilizing a graphene sheet and a flat $CaF_2$ substrate, was based on capillary condensation,[34] which can be tuned through well-established methods by varying relative humidity (RH).[34,42,43] We varied RH by purging the cell with $N_2$ of different RHs. The RH of the $N_2$ was tuned by changing the relative flow rate of a mixture comprising dry $N_2$ and wet $N_2$ while maintaining the overall $N_2$ flow rate into the sample cell. To guarantee the RH control, a microscopic humidity sensor was connected to the sample cell near the cell outlet to measure the RH in the sample cell (Fig. S3).

To show the tunable $h$ by varying RH, we conducted AFM measurements under controlled RH. The data measured across the 100×100 μm² region are shown in Fig. S14a. Decreases the RH from ~25% to ~5%, $h$ decreases from ~8 Å to below 3 Å gradually over approximately 6 hours (Fig. S14b). Importantly, the variation of $h$ with RH can be accurately modeled by the Kelvin equation (Eq. S13, see Fig. S14c), confirming that the effective confinement of water between the graphene sheet and the $CaF_2$ substrate predominantly arises from capillary condensation.[34,42,44] In our measurements, we limited the RH in the low range (~40% to ~5%) to avoid bulk water absorption at the graphene/air interface.

$$h = \frac{-2\sigma \cos\theta}{k_B T \rho_N \ln(\text{RH})}, \quad \text{(S13)}$$

where $\sigma$ is the surface tension of water at room temperature $T$,[45] $k_B$ is the Boltzmann constant and $\rho_N$ is the number density of water, $\theta$ is the contact angle of water on the walls' material. This equation has been found to hold tor both strongly hydrophilic (such as mica) and weakly hydrophilic (like graphite) capillaries, even when subjected to atomic-scale confinement.[34] The effective water contact angle of a $CaF_2$/graphene system is in the range of 40° to 60°.[46] Remarkably, within experimental uncertainty, the calculated $h$ as a function of RH using Eq. S13 provides quantitative agreement with the experimental data, underscoring that capillary condensation predominantly accounts for the successful confinement of water.

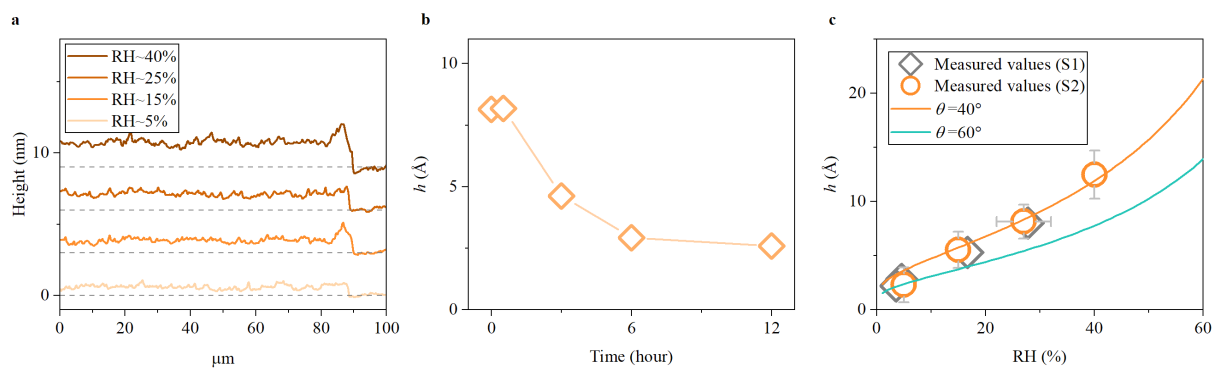

**Fig. S14 | Tuning the thickness of the nanoconfined water. a**. AFM height profiles probing the graphene edge at different RHs. For each RH, the sample was allowed to equilibrate for more than 6 hours before measurements. The data are offset by 3 nm for clarity and the dashed lines serve as zero lines. **b**. $h$ as a function of the time upon the RH change from 25% to 5%. **c**. $h$ as a function of the RH. The lines show the predictions by the Kelvin equation for two $CaF_2$/graphene water contact angles. In (**b**) and (**c**), $h$ was calculated from the AFM height profiles assuming the graphene thickness of 3.3 Å.

## S6. Amplitude Calibration of the SFG Spectra at *ssp* Polarization

The nanoconfined water sample HD-SFG spectra at *ssp* polarization were normalized with that from the gold at *ssp* polarization according to Eq. S4. To obtain the absolute value of the effective surface non-linear susceptibilities ($\chi^{(2)}_{ssp,\text{eff,sample}}$), one needs to calibrate the amplitude of the normalized spectra ($\chi^{(2)}_{ssp,\text{measured}}$) via:

$$\chi^{(2)}_{ssp,\text{eff,sample}} = \chi^{(2)}_{ssp,\text{measured}} \frac{r^*_{\text{CaF}_2/\text{gold}} \chi^{(2)}_{ssp,\text{eff,CaF}_2/\text{gold}}}{r^*_{\text{sample}}}, \quad (S14)$$

where $\chi^{(2)}_{ssp,\text{eff,CaF}_2/\text{gold}}$ is the effective SFG signal at the CaF$_2$/gold interface, $r^*_{\text{CaF}_2/\text{gold}}$ and $r^*_{\text{sample}}$ are the reflectivity coefficients for the *s*-polarized LO beams at the CaF$_2$/gold interface and CaF$_2$/water/graphene interface, respectively. The reflectivity coefficients can be calculated via Eq. S15. In this study, the CaF$_2$/water/graphene interface was regarded as a CaF$_2$/air interface for the calculation (see Section S7 for details).

$$r = \frac{n_i \cos\theta_i - n_j \cos\theta_j}{n_i \cos\theta_i + n_j \cos\theta_j}, \quad (S15)$$

where $\theta_i$ and $\theta_j$ are the refracted angle of corresponding light (SF, vis, IR) in bulk medium *i* and *j*. $n_i$ and $n_j$ are the refractive index of corresponding light in bulk medium *i* and *j*.

To get $\chi^{(2)}_{ssp,\text{eff,sample}}$, one needs to know the result of $r^*_{\text{CaF}_2/\text{gold}} \chi^{(2)}_{ssp,\text{eff,CaF}_2/\text{gold}}$ which is an unknown value. Here, we obtained the $r^*_{\text{CaF}_2/\text{gold}} \chi^{(2)}_{ssp,\text{eff,CaF}_2/\text{gold}}$ in the following three steps:

1. To obtain the absolute value of the effective surface non-linear susceptibilities at the air/D$_2$O interface ($\chi^{(2)}_{ssp,\text{eff,air/D}_2\text{O}}$), we calibrated the amplitude of the measured normalized spectra ($\chi^{(2)}_{ssp,\text{air/D}_2\text{O}}$) via:

$$\chi^{(2)}_{ssp,\text{eff},\text{air}/D_2O} = i\chi^{(2)}_{ssp,\text{air}/D_2O} \frac{r^*_{\text{air}/zqz}\left|\chi^{(2)}_{ssp,\text{eff},\text{air}/zqz}\right|}{r^*_{\text{air}/D_2O}}, \tag{S16}$$

where $r^*_{\text{air}/D_2O}$ and $r^*_{\text{air}/zqz}$ are the reflectivity coefficients of the *s*-polarized LO beams at the air/D$_2$O interface and the air/zqz interface, respectively. $\chi^{(2)}_{ssp,\text{air}/D_2O}$ is the measured spectrum at the air/D$_2$O interface normalized by the signal at the air/zqz interface. $\chi^{(2)}_{ssp,\text{eff},\text{air}/zqz}$ is the effective SFG signal at the air/zqz interface which is a known value and can be calculated from the parameters listed in Tables S1 and Eq. S17.[47,48] For H$_2$O, frequency-dependent refractive index was employed.[48,49]

$$\chi^{(2)}_{ssp,\text{eff},\text{air}/zqz} = 2L_{yy}(\omega_{\text{SF}})L_{yy}(\omega_{\text{vis}})L_{xx}(\omega_{\text{IR}})\cos\theta_i(\omega_{\text{IR}})\chi^{(2)}_q l_c, \tag{S17}$$

where $\chi^{(2)}_q \approx 8\times10^{-13}\text{mV}^{-1}$ is the second-order susceptibility of the zqz. $\omega_{\text{SF}}$, $\omega_{\text{vis}}$, and $\omega_{\text{IR}}$ are the frequency of the corresponding beam, respectively. $\theta_i$ is the incident angle of the IR beam in bulk medium $i$ as shown in Fig. S15. $L_{ii}$ ($i=x,y,z$) is the $ii$ component of the Fresnel coefficients, and is given by:

$$L_{xx} = \frac{2n_i\cos\theta_j}{n_i\cos\theta_j + n_j\cos\theta_i}, \tag{S18}$$

$$L_{yy} = \frac{2n_i\cos\theta_i}{n_i\cos\theta_i + n_j\cos\theta_j}, \tag{S19}$$

$$L_{zz} = \frac{2n_j\cos\theta_i}{n_j\cos\theta_i + n_i\cos\theta_j}\frac{n_i^2}{n'^2}, \tag{S20}$$

where $n'$ is the refractive index of the interfacial layer. Detailed discussion on $n'$ can be found in Section S7. The beam configuration of the SFG measurement is displayed in Fig. S15.

The effective coherence length for the reflected SFG, $l_c$ is calculated by:

$$l_c = \frac{1}{k_{2z}(\omega_{SF}) + k_{2z}(\omega_{vis}) + k_{2z}(\omega_{IR})} \approx 43 \text{ nm}, \tag{S21}$$

where $k_{2z}$ is the wavevector of corresponding lights in medium $j$ along the $z$-axis. The incidence angles of IR and visible beams in air are 50° and 64°, respectively. By using Eqs. S17-S21 and the parameters listed in Tables S1, we obtain $\chi^{(2)}_{ssp,\text{eff,air/zqz}} = 5.6 \times 10^{-21} \text{ m}^2\text{V}^{-1}$. From Eq. S16, we obtain $\chi^{(2)}_{ssp,\text{eff,air/D}_2\text{O}} = 7.2 \times 10^{-23} \text{ m}^2\text{V}^{-1}$.

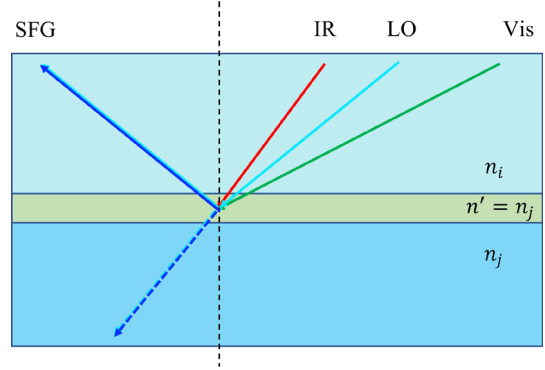

**Fig. S15 | The beam configuration of the non-colinear HD-SFG measurement.**

**Table S1. Refractive Indexes Used to Calculate the Fresnel Factors.**

| Refractive index $n$ | SF (~635 nm) | Vis (800 nm) | IR (3300 nm) |
| --- | --- | --- | --- |
| CaF$_2$ | 1.43 | 1.43 | 1.42 |
| SiO$_2$ | 1.46 | 1.45 | 1.41 |
| D$_2$O | 1.33 | 1.33 | 1.25 |
| zqz | 1.54 | 1.54 | 1.52 |

2. Assuming that $\chi^{(2)}_{yyz}$ is the same at the CaF$_2$/D$_2$O interface and at the air/D$_2$O interface, one can get:

$$\chi^{(2)}_{yyz,D_2O} = \frac{\chi^{(2)}_{ssp,\text{eff},CaF_2/D_2O}}{F^{CaF_2/D_2O}} = \frac{\chi^{(2)}_{ssp,\text{eff},air/D_2O}}{F^{air/D_2O}}, \tag{S22}$$

where $\chi^{(2)}_{ssp,\text{eff},CaF_2/D_2O}$ is the effective nonresonant SFG signal from the CaF$_2$/D$_2$O interface. $F$ is the Fresnel factor and is given by Eq. S23. Using Eqs. S22-S23, we obtain $\chi^{(2)}_{ssp,\text{eff},CaF_2/D_2O} = 2.7 \times 10^{-22}$ m$^2$V$^{-1}$. We emphasize that the assumption of a constant $\chi^{(2)}_{yyz,D_2O}$ at the two interfaces are valid because the nonresonant response mainly has a quadrupolar origin, which is insensitive to the refractive index of the bulk medium $i$ and bulk medium $j$.[50,51] Indeed, the $\chi^{(2)}_{yyz,D_2O}$ signals measured at the CaF$_2$/D$_2$O and SiO$_2$/D$_2$O interfaces show no difference within experimental uncertainty, despite different refractive index of the CaF$_2$ and SiO$_2$ substrate, further supporting the validity of the assumption (Fig. S16a).

$$F = L_{yy}(\omega_{SF})L_{yy}(\omega_{vis})L_{zz}(\omega_{IR})\sin\theta_i(\omega_{IR}). \tag{S23}$$

3. To obtain the absolute value of the effective surface non-linear susceptibilities at the CaF$_2$/gold interface ($\chi^{(2)}_{ssp,\text{eff},CaF_2/gold}$), we calibrate the amplitude of the measured normalized spectra ($\chi^{(2)}_{ssp,CaF_2/D_2O}$) via:

$$\chi^{(2)}_{ssp,\text{eff},CaF_2/gold} = \frac{r^*_{CaF_2/D_2O}\chi^{(2)}_{ssp,\text{eff},CaF_2/D_2O}}{r^*_{CaF_2/gold}\chi^{(2)}_{ssp,CaF_2/D_2O}}, \tag{S24}$$

where $r^*_{CaF_2/D_2O}$ is the reflectivity of LO at the CaF$_2$/D$_2$O interface. $\chi^{(2)}_{ssp,CaF_2/D_2O}$ is the measured spectrum at the CaF$_2$/D$_2$O interface normalized by the signal at the CaF$_2$/gold interface. From these three steps, we finally obtain:

$$r^*_{CaF_2/gold}\chi^{(2)}_{ssp,CaF_2/D_2O} = 3.2 \times 10^{-20}\text{ m}^2\text{V}^{-1}. \tag{S25}$$

## S7. Extraction of $\chi^{(2)}_{yyz}$ from $\chi^{(2)}_{ssp,eff}$

The effective SFG signal ($\chi^{(2)}_{ssp,\text{eff}}$) at the *ssp* polarization is given by:

$$\chi^{(2)}_{ssp,\text{eff}} = F\chi^{(2)}_{yyz}. \tag{S26}$$

Note that the interfacial dielectric constant ($\varepsilon' = n'^2$) is critical to do the Fresnel factor corrections and to extract the amplitude of the *yyz* components of the measured $\chi^{(2)}$ spectra ($\chi^{(2)}_{yyz}$). Two homogeneous interfacial dielectric constant models, the Lorentz model ($n' = n_j$) and the Slab model ($n' = \sqrt{\frac{n_i^2(n_i^2+5)}{4n_i^2+2}}$), are usually employed.[47,52] Such corrections and underlying assumptions of the interfacial dielectric constant[48,53] critically affect the SFG amplitude. For the *ssp* polarization combination, our recent studies show that the Lorentz model allows for more accurate Fresnel factor corrections at aqueous interfaces,[48] while the slab model underestimates the H-bonded O-H stretch signal. Therefore, in this work, the Lorentz model is employed to do the Fresnel factor corrections at all different sample geometries.

For the air/suspended graphene/water interface, we consider it as the air/water interface to do the Fresnel factor corrections because the monolayer graphene with one-atom thickness hardly affects the corrections.[54] The interface layer for the nanoconfined water sample is intricate, comprising two interfaces: CaF$_2$/confined water and confined water/graphene interfaces, which raises the question of whether one homogeneous interfacial dielectric constant can accurately capture the interface layer. To address this question, we measured the conventional SFG signals ($|\chi^{(2)}_{ssp,\text{eff},\text{CaF}_2/\text{water}/\text{graphene}}|^2$) at the CaF$_2$/confined water/graphene interface through normalization with CaF$_2$/gold signal, and compared it with the signal ($|\chi^{(2)}_{ssp,\text{eff},\text{graphene}/\text{water}/\text{CaF}_2}|^2$) obtained from the same sample but with graphene positioned on top and normalization with air/gold signal. In both sample geometries, the spectral lineshape remains consistent. After eliminating the Fresnel factor using Eqs. S23 and S26 by considering a CaF$_2$/air interface and air/CaF$_2$ interface, respectively, the spectral amplitudes are also consistent (Fig. S16b). These results indicate that one homogeneous interfacial dielectric constant is sufficient to capture the sub-nanometer thickness interface layer, and the Lorentz model is accurate for the Fresnel factor corrections for the nanoconfined water samples.

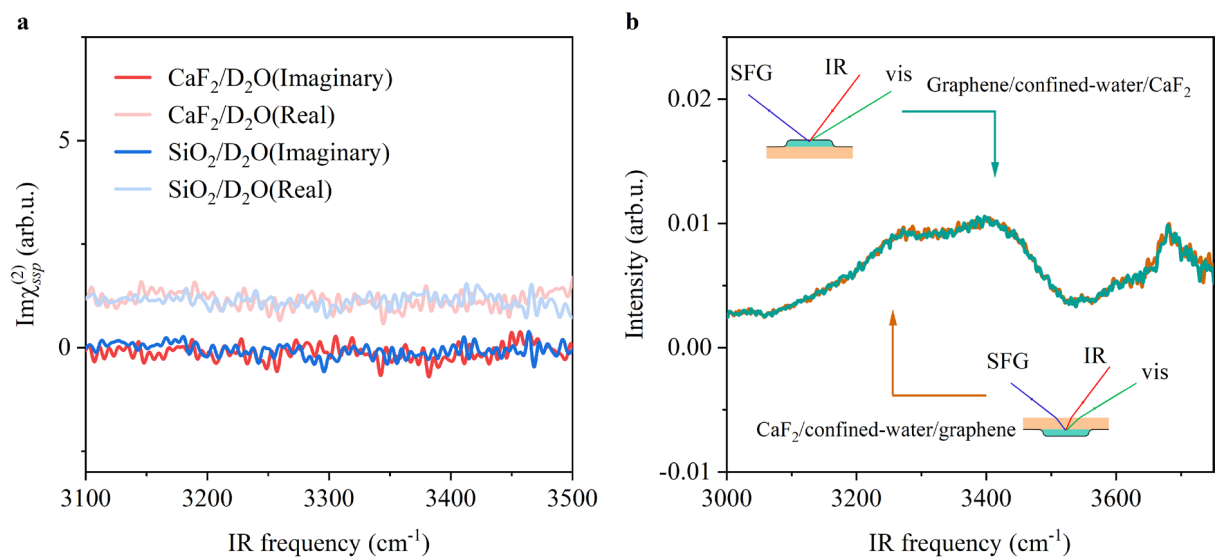

**Fig. S16 | The feasibility of the method for extracting $\chi^{(2)}_{yyz}$ from $\chi^{(2)}_{ssp}$. a.** $\chi^{(2)}_{ssp,D_2O}$ spectra measured at the CaF$_2$/D$_2$O and SiO$_2$/D$_2$O interfaces. The dashed line serves as the zero line. **b.** Conventional SFG signals at the CaF$_2$/confined-water/graphene interface and the graphene/confined-water/CaF$_2$ interface. The insets show the corresponding sample geometries and (SFG, vis, IR) beam geometries for the conventional SFG measurements.

## S8. Real part of $\chi^{(2)}_{yyz}$

Fig. S17 shows the real part of the $\chi^{(2)}_{yyz}$ spectra (Re$\chi^{(2)}_{yyz}$) of the nanoconfined three-layer water and its comparison with the sum of the water/graphene and CaF$_2$/water signals. Still, the Eq. (1) holds.

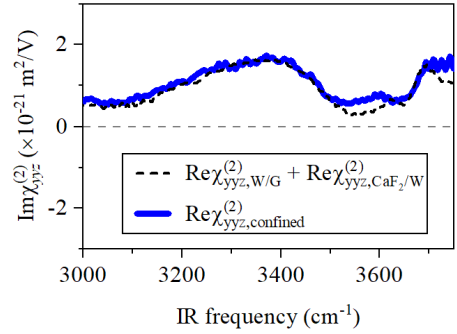

**Fig. S17 | Experimental Re$\chi^{(2)}_{yyz}$ of nanoconfined three-layer water (blue solid line).** The sum of the water/graphene and CaF$_2$/water signals obtained from HD-SFG experiments is also shown for comparison (black dashed line). A constant non-resonance contribution is subtracted in the summed spectrum because the non-resonance contribution is counted twice in the summed spectrum. Dashed grey line indicates a zero line.

## S9. Structural Information of Water Molecules Computed from AIMD Data

To examine whether the structure of the interfacial water is similar between the bulk water/graphene interface or the $CaF_2$/bulk water interface and the nanoconfined systems, we characterized the structure of the interfacial water in the following two ways; the joint probability of the two O-H group angles of water molecules and the depth profiles of the dipole orientation of water molecules. First, we calculated the joint probability distributions $P(\cos\theta_{OH_1}; \cos\theta_{OH_2})$ for the orientations of a water molecule's two O-H bond vectors,[55] where $\theta_{OH_1}$ is the angle formed by one O-H bond of a water molecule and the surface normal (z-axis) and $\theta_{OH_2}$ is the angle formed by the other O-H bond of the water molecule and z-axis. The data for the topmost water layer near the graphene sheet and near the $CaF_2$ surface are shown in Fig. S18 and S19, respectively. Fig. S18 shows that the angle distributions of the water are similar between the bulk water/graphene system and the nanoconfined water systems. A similar observation is obtained for the data near the $CaF_2$ surface (Fig. S19). This further confirms that the structure of the ~1 nm interfacial water systems and nanoconfined water systems closely resemble, in good agreement with the conclusion obtained from the SFG spectra. Furthermore, we computed the dipole orientation profiles. The data shown in Fig. S20 further confirms the notation that the interfacial water system and nanoconfined water systems are similar.

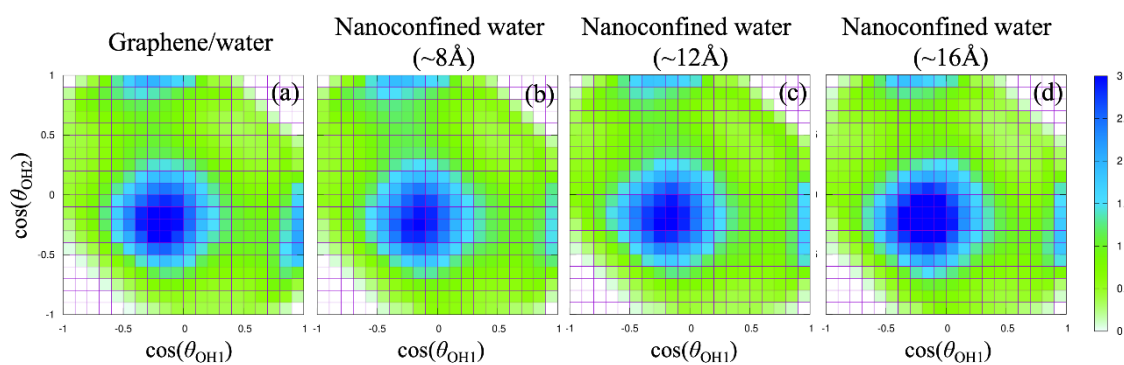

**Fig. S18 | Joint probability distributions $P(\cos\theta_{OH_1}; \cos\theta_{OH_2})$ for the orientations of the topmost layer of water molecule's two OH bond vectors near the graphene sheet computed from the AIMD data. A**. The water/graphene interface system. **b-d**, the nanoconfined systems with water thickness of ~ 8 Å (**b**), ~ 12 Å (**c**), and ~ 16 Å (**d**). The topmost layer was defined as 0 Å < z < 4.7 Å, according to the density

minima of the number density of water shown in Fig. S4. The origin point of the z-axis was the position of the graphene sheet. All the plots are normalized by the probability distribution for the corresponding isotopic bulk environment.

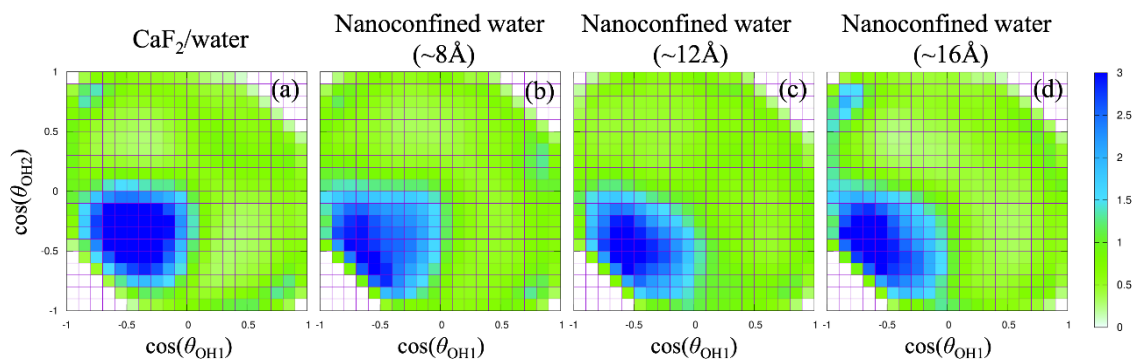

**Fig. S19 | Joint probability distributions $P(\cos\theta_{OH_1}; \cos\theta_{OH_2})$ for the orientations of the topmost layer of water molecule's two OH bond vectors near the $CaF_2$ surface computed from the AIMD data. a**. The $CaF_2$/water interface system. **b-d**, the nanoconfined systems with water thickness of ~ 8 Å (**b**), ~ 12 Å (**c**), and ~ 16 Å (**d**). The topmost layer for the $CaF_2$/water interface system was defined as -3.5 Å < z' < 0 Å while the topmost layers were defined as 7.3 Å < z < 10.8 Å, 11.3 Å < z < 14.8 Å and 15.3 Å < z < 18.8 Å for the nanoconfined water systems with their thickness of ~8 Å, ~12 Å, and ~16 Å, respectively. z (z')-axis is the surface normal whose origin point is at the graphene position (first layer F atom position), according to the density minima of the number density of water shown in Fig. S4. All the plots are normalized by the probability distribution for the corresponding isotopic bulk environment.

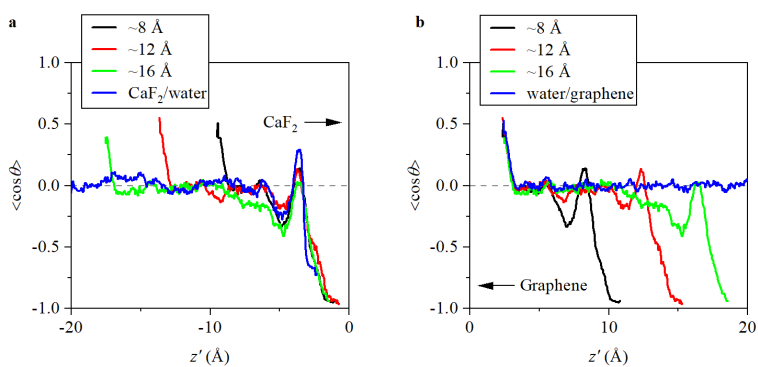

**Fig. S20 | Depth profiles of the dipole orientation of water computed by using AIMD data. a**, the $CaF_2$ surface side. **b**, the graphene sheet side. The angle $\theta$ is defined as the angle between the bisector of a water molecule and the surface normal (z- or z'-axis).

## S10. Solubility of LiCl Ions under Nanoconfinement

Upon decreasing the thickness of the nanoconfined water, the ion concentration in the confinement may increase by 1-2 fold but it is still below the solubility limit of LiCl (~18 M). To confirm the notion and show that the in-plane distribution of salt concentration between the graphene and the $CaF_2$ substrate remains uniform upon decreasing the RH, we conducted AFM measurements at different RHs. The AFM height images of the nanoconfined water between the graphene sheet and $CaF_2$ substrate are presented in Figs. S21a and b. As the RH decreases from ~25% to ~5%, the surface roughness of the sample shows only weak change, measuring 1.2 Å and 1.3 Å, and no significant precipitation of LiCl is observed. This confirms that the confined LiCl concentration remains below the solubility limit of LiCl, consistent with our estimation that the LiCl concentration of the ~8 Å confined water is around 2-3 M (see Section S1), which may increase to ~6-7 M as the RH decreases to ~5%. Furthermore, we increase the RH back to ~25%, confirming that the $Im\chi^{(2)}_{yyz}$ spectrum is reversible (Fig. S21c). This observation provides additional evidence supporting our statement.

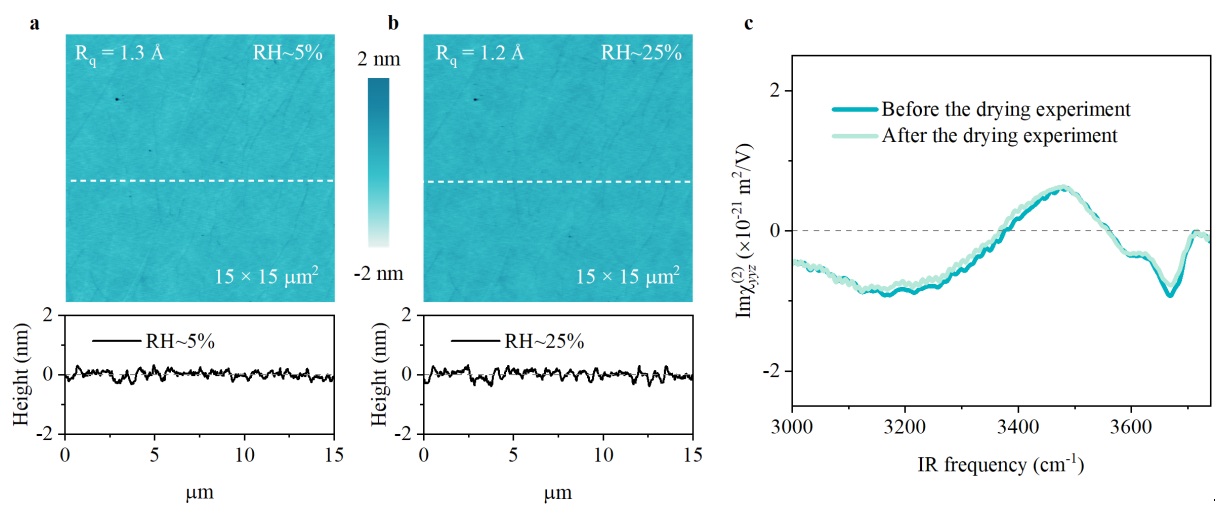

**Fig. S21 | In-plane distribution of salt concentration. a**, **b**. AFM height image of the nanoconfined water at (**a**) RH~5% and (**b**) RH~25%. The bottom panel in each AFM height image shows the typical height profiles along the white dashed lines in the corresponding AFM height image. **c**. Reversible experimental $Im\chi^{(2)}_{yyz}$ spectra of the nanoconfined water upon RH change. The dashed line serves as a zero line. The two spectra were measured at RH~25%.

## S11. Dependence of Water Thickness in Nanoconfined System on SFG Spectra

We further examined the dependence of the water thickness confined by the graphene sheet and the $CaF_2$ substrate on the SFG spectra. To this end, we computed the SFG signals with different fractions of water in the nanoconfined systems. The data are shown in Fig. S22. One can see that the variation of the SFG signal is rather insensitive to the fraction of the water molecules, consistent with the experimental data, and is also consistent with the previous theoretical prediction.[56] This insensitivity of the SFG signal to the water thickness manifests that the thickness of the water layer more than ~8 Å does not affect the SFG signal for the high ion concentration even though the $CaF_2$ surface is charged. In fact, the Debye length for the 1 M ion concentration is ~3 Å, consistent with our observation.

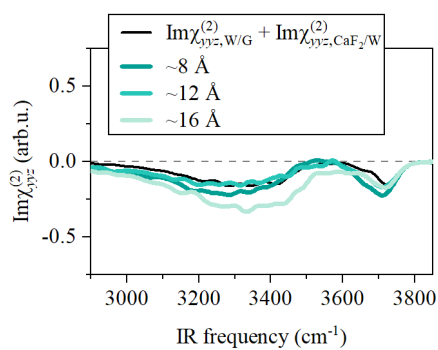

**Fig. S22 | Simulated $Im\chi^{(2)}_{yyz}$ spectra of nanoconfined water of different thicknesses computed from AIMD data.** The dashed line indicates the zero line.

## S12. SFG Spectra Computed from AIMD Data

Here, we showed the series of SFG spectra computed from the AIMD trajectories. The snapshot is shown in Fig. S23a, while the computed spectra are shown in Figs S23b-d for the confined water system, water/graphene interface, and $CaF_2$/water interface, respectively. Since the AIMD trajectories are much shorter than the MLFF-MD trajectories, the s/n ratio is worse in the spectra computed from the AIMD trajectory than those computed from the MLFF-MD trajectories for the $CaF_2$/water interface and the nanoconfined water system. Nevertheless, within the noise of the spectra, Eq. (1) in the main text is valid, meaning that the nanoconfined SFG spectrum is composed of the SFG spectra at the $CaF_2$/water interface and the water/graphene interface.

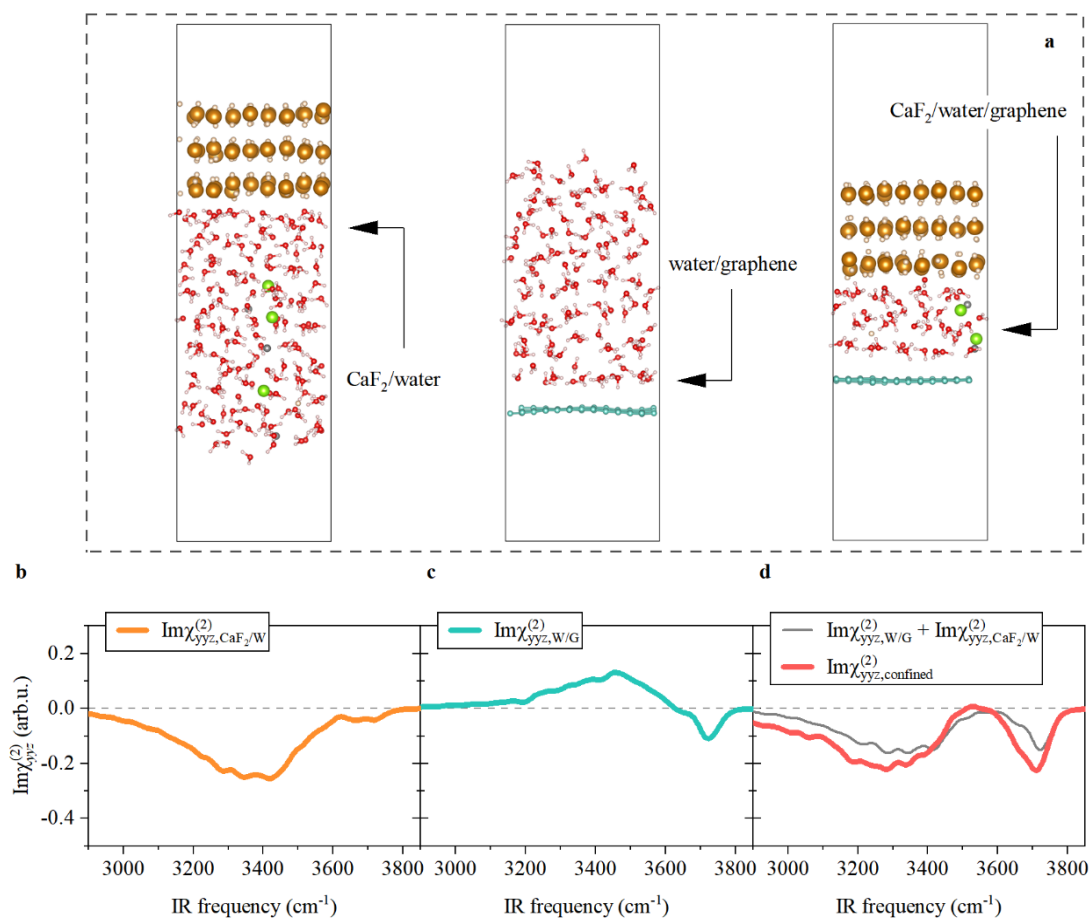

**Fig. S23 | SFG spectra calculated from AIMD. a**. Snapshots of the $CaF_2$/water interface, water/graphene interface, and nanoconfined water obtained from the AIMD simulation. The yellow, light yellow, red, light pink, green, grey, and cyan spheres indicate the Ca, F, O, H, Cl, Li, and C atoms, respectively. **b**, **c**, **d**. AIMD

based Im$\chi_{yyz}^{(2)}$ spectra of (**b**) the CaF$_2$/water interface; (**c**) the water/graphene interface; and (**d**) nanoconfined water (red line). The sum of the CaF$_2$/water (**b**) and water/graphene (**c**) SFG signals is shown in (**d**) for comparison (black line). Dashed lines in (**b-d**) indicate zero lines.

## S13. SFG Spectra Computed from AIMD and MLFF-MD Trajectories

To check the reproducibility of the MLFF-MD data, we performed the MLFF-MD at the $CaF_2$(111)/water interface for the cell with its size of $\vec{a}$, $\vec{b}$, and $\vec{c}$. The composition of the system is the same between the AIMD and MLFF-MD. The resulting depth profile of the dipole orientation of water and SFG spectra of water are shown in Fig. S24. This figure displays good agreement between the AIMD and MLFF-MD data, manifesting that the MLFF-MD can capture the trend of the AIMD data. With the current ion concentration in our study, as the ions may screen the long-range Coulomb interactions, the long-range effects may not be important. Therefore, our MLFF-MD spectra data without long-range Coulomb interactions agrees with AIMD data, this is in line with previous studies, where they did not find significant changes in spectral properties due to the long-range corrections.[57–59]

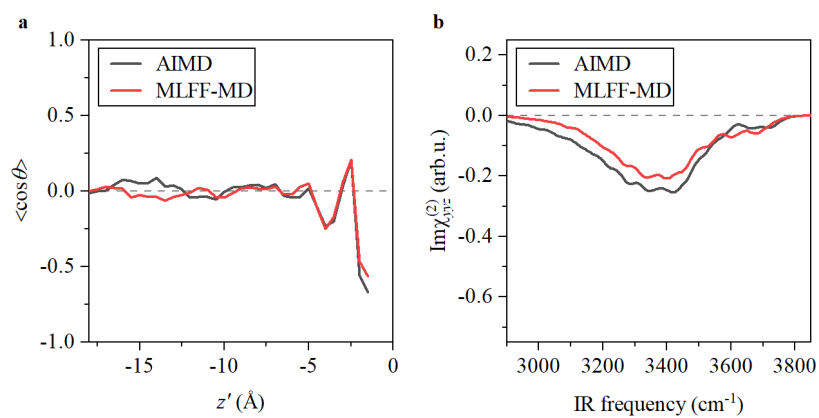

**Fig. S24 | Comparison of the AIMD data and MLFF-MD data at the $CaF_2$/water interface generated by the AIMD trajectories and MLFF-MD trajectories. a, b.** Dipole orientation (**a**) of water along the surface normal and simulated $Im\chi^{(2)}_{yyz}$ spectra (**b**) of water.

## S14. The Ions Distributions of the Nanoconfined System from MLFF-MD Trajectories

The ion distribution can give insights about the confined water structure. Fig. S25 shows the ion distributions of $Li^+$ and $Cl^-$ of the nanoconfined system (~8 Å) with high (6 M) and low (2 M) ion concentrations calculated by the MLFF molecular dynamics simulation. The ions accumulate near the charged $CaF_2$ surface with their population gradually diminishing towards zero near the charged-neutral graphene side. Despite the different ion concentrations, the SFG spectra remain unaffected, as shown in Fig. S8d.

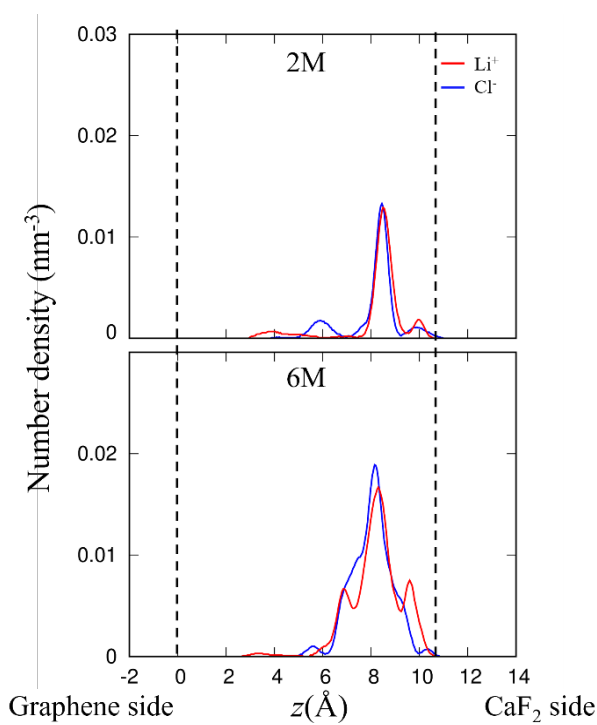

**Fig. S25 | Ion distribution along the surface normal for the nanoconfined system (~8 Å) with high (6 M) and low (2 M) ion concentrations calculated by the MLFF-MD simulation.** The dash lines marked the position of the first layer of the atoms at the graphene or $CaF_2$ sides.

# References

1. Xu, Y., Ma, Y.-B., Gu, F., Yang, S.-S. & Tian, C.-S. Structure evolution at the gate-tunable suspended graphene–water interface. *Nature* **621**, 506–510 (2023).

2. Yang, S., Zhao, X., Lu, Y. H., Barnard, E. S., Yang, P., Baskin, A., ... & Salmeron, M. Nature of the Electrical Double Layer on Suspended Graphene Electrodes. *J. Am. Chem. Soc.* **144**, 13327–13333 (2022).

3. Wang, Y., Tang, F., Yu, X., Ohto, T., Nagata, Y., & Bonn, M. Heterodyne-Detected Sum-Frequency Generation Vibrational Spectroscopy Reveals Aqueous Molecular Structure at the Suspended Graphene/Water Interface. *Angew. Chem. Int. Ed.* **n/a**, e202319503 (2024).

4. Wang, Y., Seki, T., Yu, X., Yu, C. C., Chiang, K. Y., Domke, K. F., ... & Bonn, M. Chemistry governs water organization at a graphene electrode. *Nature* **615**, E1–E2 (2023).

5. Wang, Y., Seki, T., Liu, X., Yu, X., Yu, C. C., Domke, K. F., ... & Bonn, M. Direct Probe of Electrochemical Pseudocapacitive pH Jump at a Graphene Electrode**. *Angew. Chem. Int. Ed.* **62**, e202216604 (2023).

6. Takeshita, N., Okuno, M. & Ishibashi, T. Molecular conformation of DPPC phospholipid Langmuir and Langmuir–Blodgett monolayers studied by heterodyne-detected vibrational sum frequency generation spectroscopy. *Phys. Chem. Chem. Phys.* **19**, 2060–2066 (2017).

7. Nihonyanagi, S. (二本柳聡史), Yamaguchi, S. (山口祥一) & Tahara, T. (田原太平). Direct evidence for orientational flip-flop of water molecules at charged interfaces: A heterodyne-detected vibrational sum frequency generation study. *J. Chem. Phys.* **130**, 204704 (2009).

8. Adhikari, A. Accurate determination of complex $\chi^{(2)}$ spectrum of the air/water interface. *J. Chem. Phys.* **143**, 124707 (2015).

9. Kühne, T. D., Iannuzzi, M., Del Ben, M., Rybkin, V. V., Seewald, P., Stein, F., ... & Hutter, J. CP2K: An electronic structure and molecular dynamics software package - Quickstep: Efficient and accurate electronic structure calculations. *J. Chem. Phys.* **152**, 194103 (2020).

10. Hutter, J., Iannuzzi, M., Schiffmann, F. & VandeVondele, J. cp2k: atomistic simulations of condensed matter systems. *Wiley Interdiscip. Rev. Comput. Mol. Sci.* **4**, 15–25 (2014).

11. Puchin, V. E., Puchina, A. V., Huisinga, M. & Reichling, M. Theoretical modelling of steps on the CaF2(111) surface. *J. Phys.: Condens. Matter* **13**, 2081 (2001).

12. Perdew, J. P., Burke, K. & Ernzerhof, M. Generalized Gradient Approximation Made Simple. *Phys. Rev. Lett.* **77**, 3865–3868 (1996).

13. Zhang, Y. & Yang, W. Comment on "Generalized gradient approximation made simple". *Phys. Rev. Lett.* **80**, 890 (1998).

14. Grimme, S., Antony, J., Ehrlich, S. & Krieg, H. A consistent and accurate ab initio parametrization of density functional dispersion correction (DFT-D) for the 94 elements H-Pu. *J. Chem. Phys.* **132**, 154104 (2010).

15. Goedecker, S., Teter, M. & Hutter, J. Separable dual-space Gaussian pseudopotentials. *Phys. Rev. B* **54**, 1703 (1996).

16. Krack, M. Pseudopotentials for H to Kr optimized for gradient-corrected exchange-correlation functionals. *Theor. Chem. Acc.* **114**, 145–152 (2005).

17. Bussi, G., Donadio, D. & Parrinello, M. Canonical sampling through velocity rescaling. *J. Chem. Phys.* **126**, 014101 (2007).

18. Bengtsson, L. Dipole correction for surface supercell calculations. *Phys. Rev. B* **59**, 12301–12304 (1999).

19. Zhang, L., Han, J., Wang, H., Car, R. & E, W. Deep Potential Molecular Dynamics: A Scalable Model with the Accuracy of Quantum Mechanics. *Phys. Rev. Lett.* **120**, 143001 (2018).

20. Wang, H., Zhang, L., Han, J. & E, W. DeePMD-kit: A deep learning package for many-body potential energy representation and molecular dynamics. *Comput. Phys. Commun.* **228**, 178–184 (2018).

21. Zeng, J., Zhang, D., Lu, D., Mo, P., Li, Z., Chen, Y., ... & Wang, H. DeePMD-kit v2: A software package for deep potential models. *J. Chem. Phys.* **159**, 054801 (2023).

22. Zhang, L., Lin, D.-Y., Wang, H., Car, R. & E, W. Active learning of uniformly accurate interatomic potentials for materials simulation. *Phys. Rev. Mater.* **3**, 023804 (2019).

23. Thompson, A. P., Aktulga, H. M., Berger, R., Bolintineanu, D. S., Brown, W. M., Crozier, P. S., ... & Plimpton, S. J. LAMMPS - a flexible simulation tool for particle-based materials modeling at the atomic, meso, and continuum scales. *Comput. Phys. Commun.* **271**, 108171 (2022).

24. Ohto, T., Usui, K., Hasegawa, T., Bonn, M. & Nagata, Y. Toward ab initio molecular dynamics modeling for sum-frequency generation spectra; an efficient algorithm based on surface-specific velocity-velocity correlation function. *J. Chem. Phys.* **143**, 124702 (2015).

25. Auer, B. M. & Skinner, J. L. IR and Raman spectra of liquid water: Theory and interpretation. *J. Chem. Phys.* **128**, 224511 (2008).

26. Corcelli, S. A. & Skinner, J. L. Infrared and Raman Line Shapes of Dilute HOD in Liquid H2O and D2O from 10 to 90 °C. *J. Phys. Chem. A* **109**, 6154–6165 (2005).

27. Berens, P. H. & Wilson, K. R. Molecular dynamics and spectra. I. Diatomic rotation and vibration. *J. Chem. Phys.* **74**, 4872–4882 (1981).

28. Reddy, S. K., Thiraux, R., Rudd, B. A. W., Lin, L., Adel, T., Joutsuka, T., ... & Paesani, F. Bulk Contributions Modulate the Sum-Frequency Generation Spectra of Water on Model Sea-Spray Aerosols. *Chem* **4**, 1629–1644 (2018).

29. Wen, Y. C., Zha, S., Liu, X., Yang, S., Guo, P., Shi, G., ... & Tian, C. Unveiling Microscopic Structures of Charged Water Interfaces by Surface-Specific Vibrational Spectroscopy. *Phys. Rev. Lett.* **116**, 016101 (2016).

30. Ohno, P. E., Wang, H. & Geiger, F. M. Second-order spectral lineshapes from charged interfaces. *Nat Commun* **8**, 1032 (2017).

31. Schaefer, J., Gonella, G., Bonn, M. & Backus, E. H. G. Surface-specific vibrational spectroscopy of the water/silica interface: screening and interference. *Phys. Chem. Chem. Phys.* **19**, 16875–16880 (2017).

32. Wei, F., Urashima, S., Nihonyanagi, S. & Tahara, T. Elucidation of the pH-Dependent Electric Double Layer Structure at the Silica/Water Interface Using Heterodyne-Detected Vibrational Sum Frequency Generation Spectroscopy. *J. Am. Chem. Soc.* **145**, 8833–8846 (2023).

33. Hopkins, A. J., Schrödle, S. & Richmond, G. L. Specific Ion Effects of Salt Solutions at the $CaF_2$/Water Interface. *Langmuir* **26**, 10784–10790 (2010).

34. Yang, Q., Sun, P. Z., Fumagalli, L., Stebunov, Y. V., Haigh, S. J., Zhou, Z. W., ... & Geim, A. K. Capillary condensation under atomic-scale confinement. *Nature* **588**, 250–253 (2020).

35. Gopinadhan, K., Hu, S., Esfandiar, A., Lozada-Hidalgo, M., Wang, F. C., Yang, Q., ... & Geim, A. K. Complete steric exclusion of ions and proton transport through confined monolayer water. *Science* **363**, 145–148 (2019).

36. Xu, K., Cao, P. & Heath, J. R. Graphene Visualizes the First Water Adlayers on Mica at Ambient Conditions. *Science* **329**, 1188–1191 (2010).

37. Lee, J. E., Ahn, G., Shim, J., Lee, Y. S. & Ryu, S. Optical separation of mechanical strain from charge doping in graphene. *Nat. Commun.* **3**, 1024 (2012).

38. Calado, V. E., Schneider, G. F., Theulings, A. M. M. G., Dekker, C. & Vandersypen, L. M. K. Formation and control of wrinkles in graphene by the wedging transfer method. *Appl. Phys. Lett.* **101**, 103116 (2012).

39. Wang, C., Liu, Y., Lan, L. & Tan, H. Graphene wrinkling: formation, evolution and collapse. *Nanoscale* **5**, 4454–4461 (2013).

40. Vasu, K. S., Prestat, E., Abraham, J., Dix, J., Kashtiban, R. J., Beheshtian, J., ... & Nair, R. R. Van der Waals pressure and its effect on trapped interlayer molecules. *Nat. Commun.* **7**, 12168 (2016).

41. Zabel, J., Nair, R. R., Ott, A., Georgiou, T., Geim, A. K., Novoselov, K. S., & Casiraghi, C. Raman spectroscopy of graphene and bilayer under biaxial strain: bubbles and balloons. *Nano letters* **12**, 617–621 (2012).

42. Fisher, L. R., Gamble, R. A. & Middlehurst, J. The Kelvin equation and the capillary condensation of water. *Nature* **290**, 575–576 (1981).

43. Yang, G., Chai, D., Fan, Z. & Li, X. Capillary Condensation of Single- and Multicomponent Fluids in Nanopores. *Ind. Eng. Chem. Res.* **58**, 19302–19315 (2019).

44. Deroche, I., Daou, T. J., Picard, C. & Coasne, B. Reminiscent capillarity in subnanopores. *Nat Commun* **10**, 4642 (2019).

45. Song, R., Zou, T., Chen, J., Hou, X. & Han, X. Study on the Physical Properties of LiCl Solution. *IOP Conf. Ser.: Mater. Sci. Eng.* **562**, 012102 (2019).

46. Kim, D., Kim, E., Park, S., Kim, S., Min, B. K., Yoon, H. J., ... & Cho, M. Wettability of graphene and interfacial water structure. *Chem* **7**, 1602–1614 (2021).

47. Yu, C. C., Seki, T., Chiang, K. Y., Tang, F., Sun, S., Bonn, M., & Nagata, Y. Polarization-Dependent Heterodyne-Detected Sum-Frequency Generation Spectroscopy as a Tool to Explore Surface Molecular Orientation and Ångström-Scale Depth Profiling. *J. Phys. Chem. B* **126**, 6113–6124 (2022).

48. Yu, X., Chiang, K.-Y., Yu, C.-C., Bonn, M. & Nagata, Y. On the Fresnel factor correction of sum-frequency generation spectra of interfacial water. *J. Chem. Phys.* **158**, 044701 (2023).

49. Hale, G. M. & Querry, M. R. Optical Constants of Water in the 200-nm to 200-μm Wavelength Region. *Appl. Opt., AO* **12**, 555–563 (1973).

50. Yamaguchi, S. (山口祥一), Shiratori, K. (白鳥和矢), Morita, A. (森田明弘) & Tahara, T. (田原太平). Electric quadrupole contribution to the nonresonant background of sum frequency generation at air/liquid interfaces. *J. Chem. Phys.* **134**, 184705 (2011).

51. Shen, Y. R. Revisiting the basic theory of sum-frequency generation. *J. Chem. Phys.* **153**, 180901 (2020).

52. Chiang, K. Y., Seki, T., Yu, C. C., Ohto, T., Hunger, J., Bonn, M., & Nagata, Y. The dielectric function profile across the water interface through surface-specific vibrational spectroscopy and simulations. *Proc. Natl Acad. Sci. USA* **119**, e2204156119 (2022).

53. Zhuang, X., Miranda, P. B., Kim, D. & Shen, Y. R. Mapping molecular orientation and conformation at interfaces by surface nonlinear optics. *Phys. Rev. B* **59**, 12632–12640 (1999).

54. Moloney, E. G., Azam, Md. S., Cai, C. & Hore, D. K. Vibrational sum frequency spectroscopy of thin film interfaces. *Biointerphases* **17**, 051202 (2022).

55. Odendahl, N. L. & Geissler, P. L. Local Ice-like Structure at the Liquid Water Surface. *J. Am. Chem. Soc.* **144**, 11178–11188 (2022).

56. Das, B., Ruiz-Barragan, S. & Marx, D. Deciphering the Properties of Nanoconfined Aqueous Solutions by Vibrational Sum Frequency Generation Spectroscopy. *J. Phys. Chem. Lett.* **14**, 1208–1213 (2023).

57. Yue, S., Muniz, M. C., Calegari Andrade, M. F., Zhang, L., Car, R., & Panagiotopoulos, A. Z. When do short-range atomistic machine-learning models fall short? *J. Chem. Phys.* **154**, 034111 (2021).

58. Omranpour, A., De Hijes, P. M., Behler, J. & Dellago, C. Perspective: Atomistic Simulations of Water and Aqueous Systems with Machine Learning Potentials. Preprint at https://doi.org/10.48550/arXiv.2401.17875 (2024).

59. Litman, Y., Lan, J., Nagata, Y. & Wilkins, D. M. Fully First-Principles Surface Spectroscopy with Machine Learning. *J. Phys. Chem. Lett.* **14**, 8175–8182 (2023).